\documentclass[a4paper,11pt]{article}

\pdfoutput=1
\usepackage{jheppub}

\usepackage{graphicx,mathtools,caption,subcaption,footmisc,mathrsfs}
\def\fig#1{Fig.\,\ref{#1}}
\def\App#1{Appendix \ref{#1}}

\title{Thermal Product Formula for Shear Modes}

\author[a]{Jyotirmoy Bhattacharya }
\author[a]{, Nibedita Padhi }
\author[a]{, Aditya Sharma }
\author[a]{and Sourav Singha}
\affiliation[a]{Department of Physics, Indian Institute of Technology Kharagpur, Kharagpur 721302, India}

\emailAdd{jyoti@phy.iitkgp.ac.in, nibedita.phy.95@gmail.com, adsharma.d1d4@gmail.com, singhasourav000@gmail.com}

\abstract{ We investigate the validity of the thermal product formula proposed in \cite{Dodelson:2023vrw}, for the shear channel fluctuations of R-charged black branes in AdS$_5$ where the shear mode is coupled with charge diffusion mode at non-zero momentum.
When these modes are suitably decoupled, we are able to obtain an exact formula for the two point functions of the boundary current and energy-momentum tensor in terms of the quasinormal modes of this channel. This exact formula is a simple modification of the previous version of the product formula. We also obtain a similar formula for the case involving a boundary global R-symmetry anomaly, when we have a bulk Chern-Simons term which introduces additional couplings in the shear channel. Also based on insights from the quasinormal mode spectrum, we report on an instability as well as the presence of high momentum long-lived modes associated with large values of the anomaly coefficient.
}

\begin{document}

%preprint
% \begin{flushright}
% \small{preprint}
% \end{flushright}

%
\maketitle
\flushbottom
%
%%%%%%%%%%%%%%%%%%%%%%%%%%%%%

%****************************************
\section{Introduction}
%****************************************

The study of damped oscillations of small perturbation of blackholes, characterised by the quasinormal frequencies, has been an important and active area of research for more than five decades. It is well known that, for blackholes in asymptotically flat spacetimes, such quasinormal modes (QNMs) have tremendous significance in the understanding of blackhole mergers and the associated gravitational wave physics \cite{LIGOScientific:2016aoc}. For blackholes in AdS, QNMs provide enormous insights into the dynamics of thermal states of strongly coupled conformal field theories through the AdS/CFT correspondence \cite{Maldacena:1997re,Gubser:1998bc,Witten:1998qj,Witten:1998zw,Horowitz:1999jd}.
In this holographic context, the QNMs capture the process of equilibration when the blackhole is linearly perturbed. In particular,
the dispersion of lowest frequency QNMs for AdS black branes maps to the hydrodynamic dispersions of the boundary CFT. This connection between
relativistic hydrodynamics and the long-wavelength blackhole perturbations have been elegantly established even at the non-linear level \cite{Bhattacharyya:2007vjd,Baier:2007ix,Hubeny:2011hd}.
%In this paper, we are interested in QNMs of AdS black branes beyond the hydrodynamic limit.

The equations representing linearized perturbations of AdS black branes have two distinct information about the boundary CFT which may be retrieved by solving them
with two different boundary conditions. The first set of information is given by the QNMs which govern the relaxation of the system back to equilibrium. The
QNMs are computed by solving the linearized bulk equations with normalizable boundary conditions, ensuring in-going condition at the horizon of the black brane. The second set of information pertains to finite temperature linear response of the system encoded in the thermal two-point correlation functions of the boundary CFT.
The CFT operators involved in these two-point functions are those dual to the bulk fields involved in the perturbation equations.
These equations are now solved maintaining in-going condition at the horizon but retaining both normalizable as well as non-normalizable modes near the boundary.
The retarded two-point functions are hence computed from the asymptotic coefficients of the normalizable and non-normalizable modes, using the standard AdS/CFT dictionary \cite{Gubser:1998bc,Witten:1998qj,Aharony:1999ti}. Due to this intricate connection involving the same set of equations, the QNMs must appear as simple poles of
the associated holographic two-point correlators \cite{Son:2002sd}. Despite this fact, apriori it is far from clear, to what extent the QNMs can determine these two-point functions. For example, there may be non-trivial residues of the poles and the two-point functions can have zeros unrelated to the QNMs.

In a recent paper \cite{Dodelson:2023vrw}, a remarkably simple structure of the two-point correlators in terms of the QNMs was revealed. It was proved that the two-sided
Wightman function ($\mathcal W$) corresponding to the scalar field fluctuation over AdS blackhole background admits a simple product formula in terms of the QNMs ($\omega_n$),
which can be schematically written as
\begin{equation}\label{schprodform}
 \mathcal W (\omega)= \frac{\zeta}{\prod_n (1 - \omega^2 / \omega_n^2)}.
\end{equation}
Here $\zeta$ is an overall normalization constant which cannot be straightforwardly determined by the QNMs.
This structure of the formula also remains valid at finite momentum.
The imaginary part of the retarded two-point functions ($G$) is related to $\mathcal W$ through the simple relation
\begin{equation}
 \mathcal W = \frac{\text{Im} ~ G}{\sinh \left( \omega \beta /2\right)},
\end{equation}
where $\beta$ is the reciprocal temperature. Besides, without any rigorous proof but through explicit numerical computations, it was demonstrated that such a formula remains valid for fluctuations of other metric components as well. Several background geometries were considered for demonstration, such as AdS blackholes, uncharged black branes and R-charged black branes. This product formula was subsequently used to derive and verify generic constraints on the QNM spectrum arising out of OPE in CFTs.  More recently, in \cite{Grozdanov:2024wgo,Grozdanov:2025ner} the thermal product formula together with duality type relations between thermal coorelation functions have been used to constrain and relate the QNMs of different channels. Exact analytical expressions of holographic thermal correlators were also obtained in \cite{Jia:2024zes}, and certain quantization conditions of the QNMs were established.

In this paper, pursuing the methodology \cite{Dodelson:2023vrw} we extend this line of investigation for a case where the fluctuation equations of two bulk fields are coupled. More specifically, we consider the shear mode fluctuations of an electrically charged AdS$_5$ black brane in Einstein-Maxwell theory together with a Chern-Simons term involving the gauge field. Here the Einstein equations couple with the Maxwell equations at non-zero momentum, which provide a non-trivial context to test the product formula \eqref{schprodform}. We choose to work within the shear sector because only this sector receives a non-zero contribution from the Chern-Simons term, given our choice of the background. The metric fluctuations couples with the CFT energy momentum tensor operator, while the bulk $U(1)$ gauge field couples with a R-symmetry current of the boundary theory. The presence of the Chern-Simons term in the bulk implies that there is a 't Hooft's anomaly in the global R-symmetry from the boundary CFT point of view. We should point out that this system has been instrumental in the discovery of remarkable new phenomena in hydrodynamic transport related to global anomalies \cite{Bhattacharyya:2007vs,Banerjee:2008th,Erdmenger:2008rm,Son:2009tf} (also see \cite{Kharzeev:2010gr,Jensen:2012kj,Golkar:2012kb,Jensen:2013kka,Jensen:2013rga,Haehl:2013hoa,Azeyanagi:2014sna,Azeyanagi:2015gqa}).

Although the metric and the gauge field fluctuations mix with each other, fortunately it is possible to decouple them using a set of master variables \cite{Kodama:2003jz, Edalati:2010hk,Matsuo:2009xn}. We find that the thermal product formula in the form \eqref{schprodform} is valid in terms of the operators
dual to such master variables. The master equations in the shear sector has a structure similar to the equation for the scalar field and hence
the proof of \cite{Dodelson:2023vrw} straightforwardly carries over to our case. Now, the operators dual to the master variables are composites constituted from the boundary energy-momentum tensor and R-current operators. In such a situation, when we consider the analogue of \eqref{schprodform} for the relevant components of the boundary currents (collectively represented by $\mathcal J$), we find the following (perhaps expected) form of the product formula
\begin{equation}\label{curprodsch}
 \langle \mathcal J \mathcal J \rangle_ \mathcal W = \sum_i \frac{\zeta_i ~\mathscr C_i(\omega , k)}{\prod_n (1 - \omega^2 / \left(\omega^{i}_n \right)^2 )}.
\end{equation}
where $i$ runs over the number of master variables necessary for the question at hand. Here the functions $\mathscr C_i(\omega , k)$ are simple functions
of $\omega$ and $k$ which we determine using the holographic dictionary. However, the numbers $\zeta_i$ does not appear to be determined
by the QNMs. We find that, just like the QNMs, $\zeta_i$ depends on the momentum $k$ and other parameters of Lagrangian and background solution.
The general lesson here is that although the correlators of the conserved currents in a CFT may not always have a simple form like \eqref{schprodform}, it
might have a slightly more intricate but relatively simple structure like \eqref{curprodsch}.
We have performed this analysis keeping the coefficient of Chern-Simons term ($\kappa$) arbitrary.
For a specific choice of $\kappa$ our system admits an appropriate supergravity truncation, implying that our conclusions about the structure
of the product formula \eqref{curprodsch} is also valid for $\mathcal N = 4$ supersymmetric Yang-Mills theory (SYM).

While investigating the above question we stumbled upon a few interesting observations related to the QNM spectrum of our system \footnote{To the best of our knowledge, a detailed analysis of the QNM spectrum of electrically charged black branes in the presence of the Chern-Simons term, has never been performed before.}.
In the presence of the Chern-Simons term, our black brane develops a dynamical instability characterised by positive imaginary pieces of some
QNMs. We find that this instability disappears below a threshold value
of $\kappa$ which is determined by the background charge density. Also, above this threshold the instability only exists for a finite range of the
momentum $k$. Consequently, there exists a region $\mathfrak{R}$ of the parameter space involving $k$ and $\kappa$ where the instability exists.
On a slightly different note, in the extremal limit, our black brane develops a $\text{AdS}_2 \times  \mathbb R^3$ near horizon region. For small frequencies the extremal master equations resemble the equations of a minimally coupled massive scalar field in $\text{AdS}_2$. This mass for one of master variables goes below the
$\text{AdS}_2$ Breitenlohner-Freedman (BF) bound for a certain region in the parameter space involving $k$ and $\kappa$. Curiously this region has
a strong schematic resemblance with the unstable region $\mathfrak{R}$ at finite temperature. This perhaps implies that the source of this instability
lies in the near horizon dynamics.

Another interesting observation related to the QNM spectrum is  the existence of long-lived modes at sufficiently high values of
$\kappa$. If we fix the value of $\kappa$ and analyze the spectrum for various values of momentum $k$, we find that as we increase the momentum
some of the QNMs approach the real axis, i.e. their negative imaginary parts tends to zero. This implies that relaxation times corresponding to
those fluctuations become arbitrarily large. This phenomenon has been noted previously in similar systems involving magnetically charged
black branes \cite{Waeber:2024ilt, Meiring:2023wwi, Haack:2018ztx}.

Our paper is organized as follows. In section \ref{sec:eqintro}, we introduce our system along with the relevant notations.
In section \ref{sec:zerokappa}, we present a verification of the thermal product formula
for $\kappa =0$. Even in this case, there is mixing between the shear and diffusion modes where our important observations
related to the validity of the thermal product formula can be made without the intricacies related to the Chern-Simons term.
Under this section, we introduce the master variables in \ref{ssec:masvar}, we describe the recipe for numerically computing the QNMs and
the retarded correlators in \ref{subsec:QNM} and \ref{sssec:RN2pt} respectively, and in \ref{ssec:Prodform} we discuss the validity of the thermal product formula.
Finally, we conclude this section with a brief discussion of the extremal limit in \ref{ssec:RNext}.
In section \ref{sec:kappa}, we repeat the analysis for $\kappa \neq 0$. Under this section, we first present the
master variables after significantly refining the result of \cite{Matsuo:2009xn}. Subsequently, in \ref{sssec:CSQNM} the
QNM spectrum has been discussed and the details of the aforementioned instability and long-lived modes have been reported. In \ref{ssec:CS2pt},
the product formula involving QNMs is matched with a direct computation of the two-point functions.
In \ref{ssec:CSext}, we again consider the extremal limit and report on instabilities of the near horizon geometry for generic non-zero $\kappa$.
In \App{App:2pt} we provide the holographic recipe expressing the two point functions of the shear channel components of the conserved
currents  in terms of those associated with the master variable.
We believe that the final result in \App{App:2pt} is new and important, but we choose to remove
it from the main text to present our main results with greater clarity.
%
%
%

%*******************************************************************************************************
\section{Linear fluctuations in the Shear channel} \label{sec:eqintro}
%*******************************************************************************************************

We will work in 5D with the following system
\begin{equation}\label{action}
\mathcal S = \frac{1}{16 \pi G_5} \int d^5x \sqrt{-g} \left( \mathcal R + 12 - \frac{1}{4} F_{\mu \nu}F^{\mu \nu}
- \frac{\kappa}{3} \epsilon^{\alpha \mu \nu \sigma \rho} A_{\alpha} F_{\mu \nu} F_{\sigma \rho}  \right)
\end{equation}
In addition to the usual Maxwell term we also have a Chern-Simons term for the gauge field whose coefficient $\kappa$
translates to the 't Hooft's anomaly coefficient of the boundary theory \footnote{For consistent truncation of type-IIB SUGRA in $AdS_5 \times S^5$, we must have $\kappa = \frac{1}{4 \sqrt{3}}$ \cite{Behrndt:1998jd,Son:2006em}. This value corresponds to the anomaly coefficient of a particular $U(1)$ subgroup of the $SO(6)$ R-symmetry of $\mathcal N=4$ SYM. However, in our subsequent numerical analysis we will experiment with different values of $\kappa$ to highlight the effect of the Chern-Simons term in our results. Note that, if we set $\kappa =0$, this also corresponds to an appropriate SUGRA truncation for a different non-anomalous subgroup of $SO(6)$.
}.
The equations of motion which follow from this Lagrangian are given below. The Einstein
equation is given by 
\begin{equation} \label{Eineq}
\begin{split}
\mathcal R_{\mu \nu} - \frac{1}{2} \mathcal R g_{\mu \nu}- 6 g_{\mu \nu} = T_{\mu \nu} ,
\end{split}  
\end{equation} 
where the energy momentum tensor is given by 
\begin{equation}
T_{\mu \nu} = \frac{1}{2} \left( F_{\mu \alpha} F_{\nu}^{~\alpha}
- \frac{1}{4} g_{\mu \nu} F_{\alpha \beta}F^{\alpha \beta}\right)~.
\end{equation}
The Maxwell equation is given by
{\footnote{Our convention for the Levi-Civita is $\epsilon^{\nu \alpha \beta \sigma \rho} = \frac{1}{\sqrt{-g}} \hat{\epsilon}^{\nu \alpha \beta \sigma \rho}$,
where $\hat{\epsilon}^{01234} = + 1$.}}
\begin{equation} \label{MaxScaleq}
\begin{split}
& \nabla_{\mu} F^{\mu \nu} =
\kappa \epsilon^{\nu \alpha \beta \sigma \rho}  F_{\alpha \beta} F_{\sigma \rho} ~.
\end{split}  
\end{equation} 
\noindent
We intend to study shear sector fluctuations about specific backgrounds which are solutions to these set of equations. In Schwarzschild-like coordinates, we would like to make the following spherically symmetric ansatz for the background metric and gauge field
\begin{equation}\label{Schmetanza}
 \begin{split}
ds^2 &= g^{(0)}_{\mu \nu}dx^\mu dx^\nu = \frac{1}{z^2} \left(- f(z) d t^2 + \frac{d z^2}{f(z)} +
%(d {x_1}^2 + d {x_2}^2+ d {x_3}^2)
d\vec x^2 \right),\\
  A &= A^{(0)}_\mu dx^\mu = \psi(z) dt.
 \end{split}
\end{equation}
Here $z$ is the radial coordinate in AdS which goes to zero at the boundary.
In these coordinates, we choose the following gauge fixing condition
\begin{equation}\label{Schgaugecond}
g_{z\mu} =0 , ~~ A_z =0.
\end{equation}
where $\mu$ denotes the coordinates along the boundary directions.

We now consider the linearized fluctuations around the background fields
\begin{equation} \label{flucform}
\begin{split}
 g_{\mu \nu} &= g^{(0)}_{\mu \nu}(z) + \delta g_{\mu \nu} (z) ~e^{i \left( k x_1 - \omega t \right)}, \\
 A_\mu &= A^{(0)}_\mu(z) + \delta A_{\mu} (z) ~ e^{i \left( k x_1 - \omega t \right)}, \\
\end{split}
\end{equation}
where $g^{(0)}_{\mu \nu}(z)$ and $A^{(0)}_{\mu}(z)$ are the background fields in \eqref{Schmetanza}, while $\delta g_{\mu \nu}(z)$ and $\delta A_{\mu}(z)$ are the amplitudes of the small fluctuations to be treated at linear order in our analysis.
%Note that the frequency of the fluctuations $\omega$ is defined with respect to EF time coordidate $v$, and
Exploiting the rotational symmetry along the boundary spatial coordinates, we have chosen the wave-vector along the $x_1$ direction without any loss of generality. This wave vector preserves a $SO(2)$ within the $SO(3)$ representing spatial rotations on the boundary. The fluctuations in \eqref{flucform} are then classified by their transformation properties under this $SO(2)$, which we denote by $\mathcal H = \{\pm 2 , ~\pm 1 , ~ 0\}$, with each channel having a decoupled set of equations (see \cite{Jansen:2017oag} for more details).

Our interest in this paper is the sector of fluctuations with  $\mathcal H = \pm 1$, which we collectively refer to as the shear channel. If we
consider a transformation on the $x_2$ and $x_3$ coordinates given by $~ x_2 \rightarrow - x_2 , ~ \text{and}~ x_3 \rightarrow -x_3$, then
the $\mathcal H = \pm 1$ fluctuations are odd under this transformations \cite{Edalati:2010hk}.

The fluctuations of this sector are explicitly listed below
\begin{equation}\label{flucmodes}
 H_1 = {{\delta g}^{x_1}}_{x_2}, ~ F_1 = {{\delta g}^{x_2}}_{t}, ~\mathcal B_1 = \delta A_{x_2},
 ~ H_2 = {{\delta g}^{x_1}}_{x_3}, ~  F_2 = {{\delta g}^{x_3}}_{t}, ~ {\mathcal B}_2 = \delta A_{x_3}.
\end{equation}
These fluctuations do not mix with other components of the fluctuations which are  $\mathcal H = \{\pm 2 , ~ 0\}$. Hence this sector may be studied independently, switching off other fluctuations. This sector captures important transport properties in the shear and diffusion channels. For the background which we have considered here, this is the only sector that captures non-trivial effect of the Chern-Simons term, which is one of the main focus of our paper \footnote{In the charged black brane backgrounds we consider here, we only have a non-zero electric field. In such a case, the Chern-Simons term in \eqref{MaxScaleq} vanishes for the fluctuations
with $\mathcal H = \{\pm 2 , ~ 0\}$.}.
Also, the contribution from the Chern-Simons term is non-zero only for the modes with non-zero momentum $k$.
We would like to point out that if there is rotational symmetry in the $x_2 x_3$-plane then it suffices to consider only a half of these fluctuations, i.e. $ ~\delta g_{tx_2}, ~\delta g_{x_1 x_2}, ~\delta A_{x_2} $, the equation for the other half being identical. However, the Chern-Simons term which contributes to the Maxwell equation in this sector, breaks this symmetry leading to a mixing between $\delta A_{x_2}$ and $\delta A_{x_3}$ components of the fluctuations. As we shall see this leads to important physical consequences for the QNM spectrum.
%
%

%***********************************************************************************************
\section{Electrically Charged black branes ($\kappa =0$)}\label{sec:zerokappa}
% %***********************************************************************************************
%
%
%
For the R-charged black branes \footnote{ The QNMs for the 3+1D charged black branes were neatly worked out in \cite{Edalati:2010hk}. Here we repeat their analysis in 4+1D and verify the thermal product formula.} the background metric and gauge field functions \eqref{Schmetanza} are given by
\begin{equation}\label{chBH}
 \begin{split}
  f(z) &= 1 - (1+Q^2) ~z^4 + Q^2 ~z^6, \\
  \psi(z) &= \sqrt{3} Q (1-z^2).
 \end{split}
\end{equation}
The horizon is located at $z=1$ while $z \rightarrow 0$ represents the boundary.
Once we have fixed the location of the horizon, there is another free parameter $Q$ which controls both the temperature and chemical potential of the black brane,
which are respectively given by
\begin{equation}\label{Tmudef}
 T = \frac{1}{\pi} \left( 1 - \frac{Q^2}{2}\right), ~~\mu = \sqrt{3}~ Q.
\end{equation}
Hence, in the limit $Q \rightarrow \sqrt{2}$, the black brane approaches extremality.

%********************************************************************
\subsection{Decoupling of fluctuations}\label{ssec:masvar}
%********************************************************************
The equations for the fluctuations $H, F$ and $\mathcal B$ in \eqref{flucmodes}
are given by the corresponding components of the Einstein and Maxwell equations (\eqref{Eineq}, \eqref{MaxScaleq}) linearized about the charged black-brane background \eqref{chBH} (see \App{App:2pt})
{\footnote{For $\kappa =0$, these equations are identical to those of $\bar H, \bar F$ and $\bar{\mathcal B}$ which
we ignore for this discussion.}}.
One of these equations is a constraint equation. With the help of this constraint, one of the metric fluctuations can be eliminated, and the combination of the fluctuations that are invariant under the residual gauge transformation \cite{Kodama:2003jz, Edalati:2010hk} are given by
\footnote{For further details of the gauge invariant variables, decoupling and the master variables the reader is referred to the original references \cite{Kodama:2003jz, Matsuo:2009xn, Edalati:2010hk}. }
\begin{equation}\label{GIvar}
 \begin{split}
 & \mathcal G \equiv k  F_1 + \omega H_1, ~\mathcal B \equiv \mathcal B_1 ,\\
 \end{split}
\end{equation}
where $\omega, k$ are the fluctuation frequency and momentum \eqref{flucform} \footnote{It is customary to work in dimensionless parameter corresponding to $\omega$ and $k$ by appropriately dividing by suitable characteristic scale in the problem, such as the temperature or chemical potential. However we prefer to work with the dimensionful quantities without any such rescaling to avoid carrying around a cumbersome number. This will not affect any of the analysis presented here since we have fixed the horizon radius to 1 throughout, and we will report most of our results at constant $Q$. Moving to the dimensionless variables is merely an overall numerical rescaling of all the QNMs.}. The independent equations for these variables are given by
\begin{equation}\label{chBHfluc}
 \begin{split}
 & z \mathcal{G}''-3 \mathcal{G}'- \frac{\left ( f k^2 - \omega^2 \right) z \mathcal{G}}{f^2} + k z^3 \psi' \mathcal{B}'   =
   \frac{\omega^2
   f \left(z^4 \psi'^2+ 24 \left( f - 1 \right) \right) }{6 f^2 \left(k^2 f-\omega^2\right)} \left(k z^2 \psi'
   \mathcal{B} +\mathcal{G}' \right) , \\
 &  z f \left(\mathcal{B}'' + \frac{k \psi' \mathcal{G}'}{k^2
   f-\omega^2}\right)+ \left(  \frac{1}{6} z^4 \psi'^2 + 3
   f - 4 \right) \mathcal{B}'  +  z  \left(\omega^2 \left(\frac{z^2
   \psi'^2}{k^2 f-\omega^2}+\frac{1}{f}\right)  -k^2\right)\mathcal{B} = 0 ,\\
 \end{split}
\end{equation}
where, a prime denotes a derivative of the functions with respect to $z$.
Here $f(z),  \psi(z)$ are the background fields given by \eqref{chBH}. The first equation is a particular combination of the Einstein equation, while the second equation follows
from the Maxwell equation.
Following \cite{Kodama:2003jz, Edalati:2010hk} we define the master variables
\begin{equation}\label{chBHmastervar}
 \begin{split}
 \Phi_{\pm} = \left( \frac{3 k f \psi'}{2 z^2 \left(\omega^2-k^2 f\right)} \right) \mathcal G' +
 \left(\frac{\psi'^2 \left(2 k^2 f+\omega^2\right)}{2 \left( \omega^2-k^2 f \right)}+\frac{6 (1-f)}{z^4}  \mp \mathcal C \right) \mathcal B ,\\
 \end{split}
\end{equation}
where $\mathcal C$ is a constant on-shell and is given by
\begin{equation}\label{Cdef}
 \begin{split}
 \mathcal C &= \left( \frac{1}{4 z^8} \left( 3 z^4 \psi'^2 \left(-8 f+3 k^2 z^2+8\right)+z^8 \psi'^4+144 (f-1)^2 \right) \right) ^{\frac{1}{2}} \\
 & = 3  \Big( \left(3 k^2+8\right) Q^2+4 Q^4+4 \Big)^{\frac{1}{2}} ~.
 \end{split}
\end{equation}
In terms of these master variables \eqref{chBHfluc} decouples into
\begin{equation}\label{chBHmastereqn}
 \begin{split}
 \partial_z \left( \frac{f}{z} \Phi'_{\pm} \right) + \left( -\frac{4 f'}{z^2}+\frac{\omega^2}{z f}+\frac{12 f}{z^3}-\frac{k^2}{z}-\frac{12}{z^3} \pm \frac{2 z }{3} \mathcal C  \right) \Phi_{\pm}  = 0 ~.\\
 \end{split}
\end{equation}

%********************************************************************
\subsection{QNMs}\label{subsec:QNM}
\begin{figure}[h]
\centering
\begin{subfigure}{.5\textwidth}
  \centering
  \includegraphics[width=0.93\linewidth]{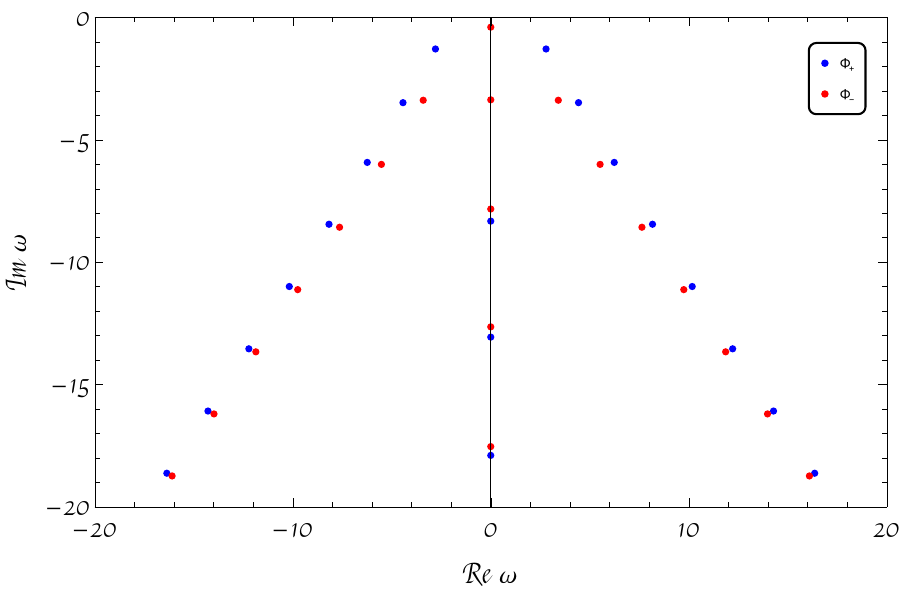}
  \caption{}\label{fig:RNAdSQNM}
\end{subfigure}%
\begin{subfigure}{.5\textwidth}
  \centering
  \includegraphics[width=0.9\linewidth]{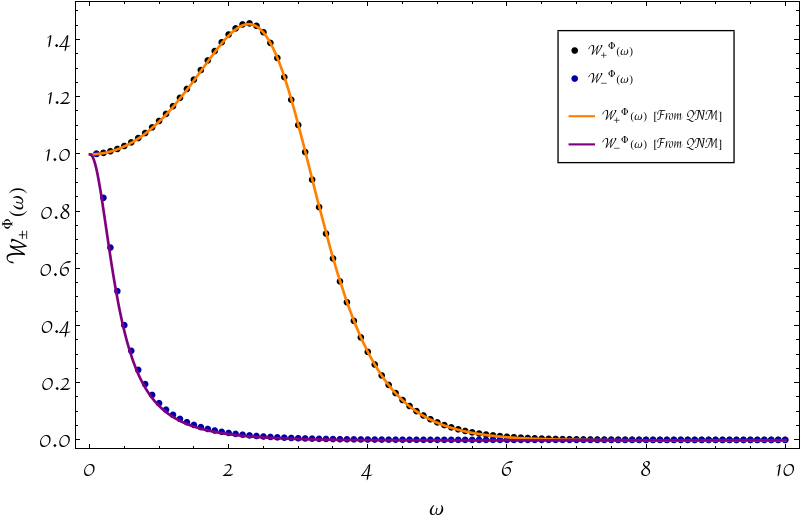}
  \caption{}\label{fig:RNAdS2pt}
\end{subfigure}
\caption{In (a), we have plotted the QNMs in the complex $\omega$ plane, which arises out of \eqref{chBHmastereqn} for $Q= 1/ \sqrt{2}$ and $k = 1.5$. In (b), we plot
the two sided Wightman function \eqref{Wdef} for the same set of parameters. The dotted lines represent a direct computation of
$\mathcal W^\Phi_{\pm}$ from a numerical solution of \eqref{chBHmastereqn} with different values of $\omega$. The solid lines overlaying the dotted lines
represent the same quantity evaluated through the product formula \eqref{RNprodform}, using the QNMs from (a). We have adjusted the normalization so that $\zeta_\pm = 1$.
We have used the tail correction formula \eqref{tailcor} with $n_0 = 15$, $r \approx 4.95$ for the on-axis poles, and $n_0 =30$, $r \approx 3.30$, $\theta \approx 0.87$ for the off-axis poles, for both $\Phi_+$ and $\Phi_-$. The excellent agreement between the dotted
and the solid lines in (b) demonstrate the validity of the formula \eqref{RNprodform}.
}
\label{fig:RNAdS}
%Mathematica file: RNAdS_Final_DataAndPlots.nb, 5D_RN_ADS_MasterEquations_Sch_Coordinates_tail_corrected_Q1byroot2
\end{figure}

%********************************************************************
The QNMs are the values of $\omega$ for which \eqref{chBHmastereqn} admits a normalizable solution which is in-going at the horizon.
The in-going boundary condition is ensured by working in a redefined master variable
\begin{equation}
\Phi_{\pm}(z) = f(z)^{- i \frac{\omega}{4\pi T}} ~z ~\tilde \Phi_{\pm}(z) ,
\end{equation}
so that $\tilde \Phi_{\pm}$ is a constant at the horizon located at $z=1$. The factor of $z$ ensures that near the boundary ($z \rightarrow 0$),
$\tilde \Phi_{\pm}$ has the following behaviour
\begin{equation}\label{mastervarbdy}
\tilde \Phi_{\pm}(z) =  \left( \frac{\mathcal S^\Phi_{\pm}}{z} + \mathcal O^\Phi_{\pm} z + \cdots  \right) ,
\end{equation}
where $\mathcal O^\Phi_{\pm}$ and $S^\Phi_{\pm}$ are respectively the normalizable and non-normalizable modes of the master field.
% % is proportional to the boundary expectation value of the operator dual to $\Phi_{\pm}$ while $\mathcal S^\Phi_{\pm}$are the corresponding sources.
Now, the QNMs are given by the discrete values of $\omega = \omega_n$ for which \eqref{chBHmastereqn} admits a regular in-going solution with $\mathcal S^\Phi_{\pm} =0$. We have obtained the values of $\omega_n$ with the help of the mathematical package `{\it{QNMspectral}}' \cite{Jansen:2017oag} {\footnote{ For using the package, including the choice of numerical parameters, we have closely followed  \cite{Dodelson:2023vrw} (see Appendix H in that paper). \label{FN:QNMspectral}}}. The QNMs so obtained has been displayed in \fig{fig:RNAdSQNM}.
%

%********************************************************************
\subsection{Two-point functions} \label{sssec:RN2pt}
%********************************************************************
%
We know that the metric and the gauge field in AdS couples with boundary energy-momentum $\mathcal T_{\mu \nu}$ and R-charge current $\mathcal J_{\mu}$ operators respectively. Our objective here is to write down a compact formula for the two point function of these operators in terms of the QNMs obtained in the previous section. But before moving on to such a thermal product formula, let us directly compute these holographic two point functions so that we are able to verify the proposed formula. In the shear sector the relevant components of these conserved currents are $\mathcal T_{t x_2}, ~ \mathcal T_{x_1 x_2}$ and $\mathcal J_{x_2}$.

In order to compute the two point function of these operators, we again solve \eqref{chBHmastereqn} but now we keep both the normalizable $\mathcal O^\Phi_{\pm}$ and non-normalizable $\mathcal S^\Phi_{\pm}$ modes, maintaining in-going boundary condition at the horizon. The in-going boundary condition translates to $\tilde \Phi_{\pm}$ being a constant at the horizon which we choose to be $1$, thus fixing an overall normalization. After numerically solving \eqref{chBHmastereqn},
we record the ratio
\begin{equation} \label{Gdef}
 G^\Phi_{\pm} (\omega, k) = \frac{\mathcal O^\Phi_{\pm}(\omega, k)}{\mathcal S^\Phi_{\pm} (\omega, k)}.
\end{equation}
Using the standard AdS-CFT dictionary, we can express (see Appendix \ref{App:2pt} for more details) the retarded two point function of the conserved currents in terms of this ratio, up to contact terms, as follows \cite{Son:2002sd, Edalati:2010hk}
\begin{equation}\label{RNallretcor}
 \begin{split}
 & \langle \mathcal T_{x_1 x_2} \mathcal T_{x_1 x_2} \rangle = \frac{\omega^2}{8 \mathcal C} \left( \mathcal C_{-} ~G^\Phi_{-} +  \mathcal C_{+} ~G^\Phi_{+} \right) ,
 ~~\langle \mathcal J_{x_2} \mathcal J_{x_2} \rangle = \frac {1}{ 2 \mathcal C }  \left( \mathcal C_{+} ~G^\Phi_{-} +  \mathcal C_{-} ~G^\Phi_{+} \right), \\
 & \langle \mathcal T_{t x_2}  \mathcal T_{t x_2} \rangle = \frac{ k^2 }{ 8 \mathcal C } \left(  \mathcal C_{-} ~ G^\Phi_{-} +  \mathcal C_{+} ~ G^\Phi_{+} \right),
 ~~\langle \mathcal T_{x_1 x_2} \mathcal J_{x_2} \rangle = \frac{3 \sqrt{3} ~k \omega Q}{4 \mathcal C}  ( G^\Phi_{+} - G^\Phi_{-} ),\\
 & \langle \mathcal T_{x_1 x_2}  \mathcal T_{t x_2} \rangle = \frac{k  \omega}{8 \mathcal C} \left(  \mathcal C_{-} ~ G^\Phi_{-}  +  \mathcal C_{+} ~ G^\Phi_{+} \right),
 ~~ \langle \mathcal J_{x_2} \mathcal T_{t x_2}  \rangle = \frac{3 \sqrt{3} ~k^2 Q}{4 \mathcal C} \left( G^\Phi_{+}-G^\Phi_{-} \right), \\
 \end{split}
\end{equation}
where $\mathcal C_\pm = -\mathcal C \pm 6 \left(1+ Q^2\right)$ and $\mathcal C$ is defined in \eqref{Cdef}.

Note that the other retarded two point functions in this sector are expressed in terms of the above up to contact terms. Hence all these retarded correlators are determined
by the two ratios \eqref{Gdef} {\footnote{Note that the ratios \eqref{Gdef} are themselves two point function of some composite operator of the conserved currents in
the shear sector.}}.

%********************************************************************
\subsection{Two-point functions from QNMs} \label{ssec:Prodform}
%********************************************************************
%
Inspired by the results in \cite{Dodelson:2023vrw}, one of our main observations in this paper is that the ratio $G^\Phi_{\pm}$, and hence all the retarded correlators in \eqref{RNallretcor} are entirely determined by the quasinormal modes obtained in \ref{subsec:QNM}. The first step is to define an analogue of the \emph{two-sided Wightman correlator} for $G^\Phi_{\pm}$ in the following way
(see Appendix A of \cite{Dodelson:2023vrw})
\begin{equation}\label{Wdef}
 \mathcal W^\Phi_{\pm}(\omega, k) = \frac{\text{Im} ~G^\Phi_{\pm} (\omega,k)}{\sinh \left( \omega \beta /2\right)}.
\end{equation}
Note that $\text{Re} ~G^\Phi_{\pm}$ is related to $\text{Im} ~G^\Phi_{\pm}$ by the usual Kramers-Kronig relation, since $G^\Phi_{\pm}$ is a retarded correlators itself.
This implies that once we know $\mathcal W^\Phi_{\pm}$, all the retarded correlators in \eqref{RNallretcor} can be straightforwardly determined.

Now the two-sided Wightman functions $\mathcal W^\Phi_{\pm}$ is expressed entirely in terms of the QNMs through the following
product formula
\begin{equation}\label{RNprodform}
 \mathcal W^\Phi_{\pm} = \zeta_\pm \underbrace{\prod_{n} \left( 1 - \frac{\omega^2}{\omega_n^2} \right)^{-1} \left( 1 - \frac{\omega^2}{{\omega^\star}_n^2} \right)^{-1}}_{\text{(off-axis poles: single line)}}      \underbrace{\prod_{n} \left( 1 - \frac{\omega^2}{\omega_n^2} \right)^{-1}}_{\text{(on-axis poles)}}
\end{equation}
where $\zeta_\pm$ are undetermined overall constants related to normalization of the operators in question.
In this product formula, the pole contributions arises from all the QNMs which can be broadly classified as on-axis (i.e. on the imaginary axis in the complex $\omega$-plane) and off-axis (i.e. away from the imaginary axis). The off-axis QNMs occur in pairs, i.e. for every $\omega_n$ there is another corresponding QNM at $(- \omega_n^\star)$. Therefore, these off-axis QNMs organize themselves into two approximate straight lines (see \fig{fig:RNAdSQNM}). The off-axis contribution to the product formula is written in such a way that we should use the values of QNMs lying on only one of these two lines. In \fig{fig:RNAdS2pt}, we have plotted $\widehat{\mathcal W}^\Phi_{\pm} = \mathcal W^\Phi_{\pm} / \zeta_\pm$ obtained from the product formula \eqref{RNprodform} using the QNMs from \fig{fig:RNAdSQNM}, and compared it with that obtained by directly solving the differential equations following the method outlined in \ref{sssec:RN2pt}.

With the help of this product formula all the correlators in \eqref{RNallretcor} can now be determined. For example, the Wightman function corresponding to the retarded correlator $\langle \mathcal T_{x_1 x_2} \mathcal T_{x_1 x_2} \rangle$ will be given by
\begin{equation}\label{TTwhit}
 \langle \mathcal T_{t x_2} \mathcal T_{t x_2} \rangle_\mathcal W = \frac{\text{Im} \langle \mathcal T_{t x_2} \mathcal T_{t x_2} \rangle}{\sinh \left( \omega \beta /2\right)}
 = \frac{ k^2 }{ 8 \mathcal C } \left(  \mathcal C_{-} ~ \zeta_- \widehat{\mathcal W}^\Phi_{-} +  \mathcal C_{+} ~\zeta_+ \widehat{\mathcal W}^\Phi_{+} \right),
\end{equation}
where $\widehat{\mathcal W}^\Phi_{\pm}$ are given entirely in terms of QNMs, while $\zeta_\pm$ remain as two undetermined constants. The constants $\zeta_\pm$ can be obtained by appropriately fitting the correlators obtained from the differential equations. The same values of $\zeta_\pm$ are applicable to all the correlators in \eqref{RNallretcor}. As an example, we have obtained the values of $\zeta_\pm$ when $Q=1$ and $k = 1.5$ and subsequently, using \eqref{TTwhit}, we have plotted
$\text{Im} \langle \mathcal T_{t x_2} \mathcal T_{t x_2} \rangle$
in \fig{fig:TTplot} (See the $\kappa = 0$ plot). In this way, the master variables help us uncover a relatively simple generalization of the product formula which is applicable for all the correlators in \eqref{RNallretcor}.

It is worthwhile to point out that the correlators \eqref{TTwhit}, along with all their counterparts which follow from \eqref{RNallretcor} collectively represented by $\mathcal W$, are expected to satisfy constraints following from unitarity and KMS conditions \cite{Dodelson:2023vrw}, which are respectively given by
\begin{equation}\label{kmsu}
 \mathcal W (\omega) \geq 0 ~ \text{for}~ \omega \in \mathbb R, ~~ \mathcal W (\omega) =  \mathcal W (-\omega).
\end{equation}
Once we use \eqref{RNprodform} into \eqref{TTwhit}, the resultant structure, together with the location of
QNMs in the complex $\omega$ plane (\fig{fig:RNAdSQNM}), ensures complete consistency with \eqref{kmsu}.

\begin{figure}[thb]
\centering
\includegraphics[width = 0.5\textwidth]{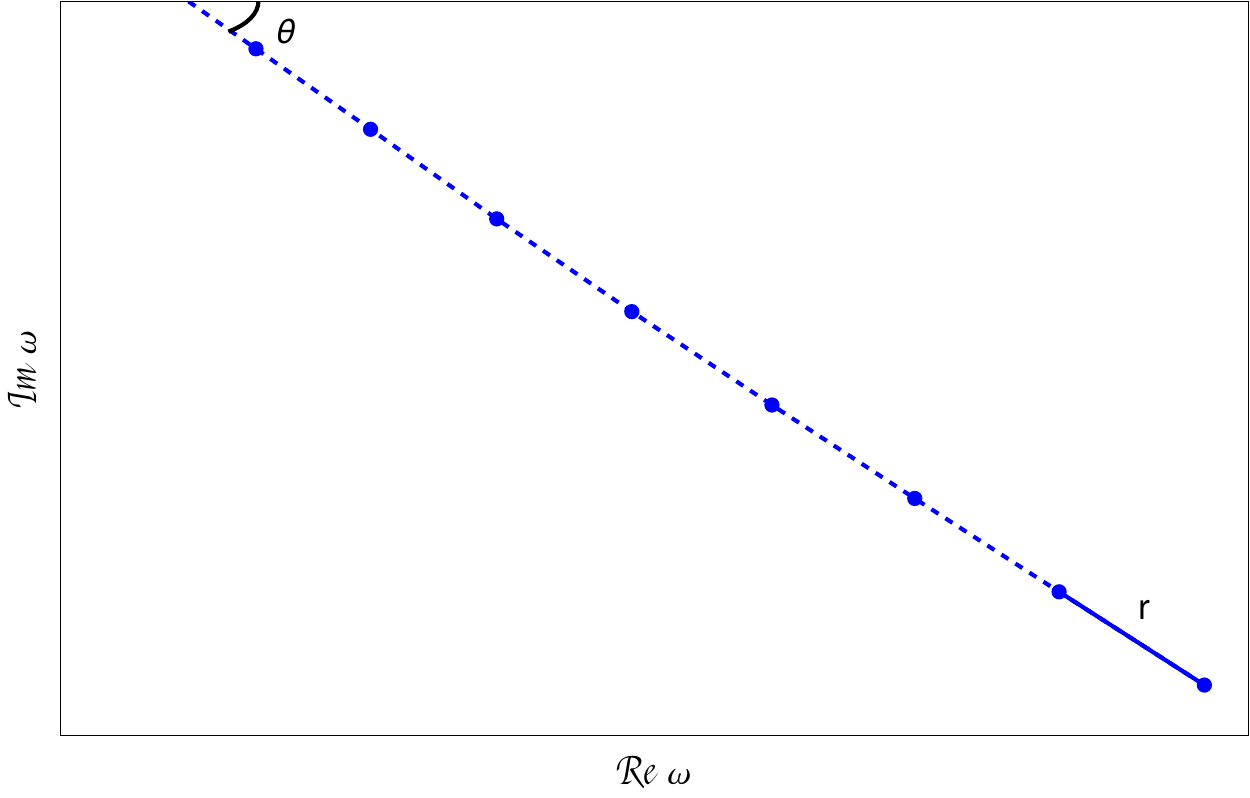}
\caption{Schematic diagram to explain the parameters $r$ and $\theta$ appearing in the formula \eqref{QNMapprox} for asymptotic QNMs.}\label{fig:SchmTCfig}
%Mathematica file: ConventionPlot.nb
\end{figure}

{\flushleft{\bf{Tail corrections:}}}
Before concluding this discussion, let us recall a very useful approximation from \cite{Dodelson:2023vrw} related to the practical application of \eqref{RNprodform}.
The product formula requires us to know all the QNMs which is not possible when we determine them numerically. In order to troubleshoot this problem, we first observe that
when the magnitude of the QNMs are very high, they organize themselves on straight lines with equal integer spacing (see \fig{fig:RNAdSQNM}).
This observation leads to the following approximate formula for QNMs with large magnitudes
\begin{equation}\label{QNMapprox}
 \omega_n = \omega_{n_0} + r e^{-i \theta} \left(n - n_{0} \right),
\end{equation}
where $n_0$ is a relatively large positive integer. In this formula, we have assumed that the QNMs which approximately lie on a single line are labeled by
integers with the one closest to the real axis corresponding to $n=1$. So, $\omega_{n_0}$ with a large integer value $n_0$ is located considerably away from the real axis.
The asymptotic formula is valid for $\omega_n$ with $n \geq n_0$ which are further away. Here $r$ denotes the distance between
these asymptotic QNMs, which is approximately constant, while $\theta$ denotes the acute angle between the line and the real axis (see the schematic diagram in \fig{fig:SchmTCfig}). For the on-axis QNMs, we should use $\theta = \pi /2$ in \eqref{QNMapprox}.
Clearly, higher the value of $n_0$, the better is the approximation.

We can now obtain the QNMs numerically up to some value of $n_0$ to be used in \eqref{RNprodform}, while for the rest of the QNMs we can use
\eqref{QNMapprox} converting the product over poles into the following tail correction formula
{\small {
\begin{equation}\label{tailcor}
 \begin{split}
 & \text{Off-axis:}~
 \prod_{n=n_0 + 1}^\infty \left( 1 - \frac{\omega^2}{\omega_n^2} \right)^{-1} \left( 1 - \frac{\omega^2}{{\omega^\star}_n^2} \right)^{-1} \\
 & = \frac{ \Gamma \left(1 + \frac{e^{i \theta }}{r} (\omega_{n_0}-\omega)  \right)
\Gamma \left(1 + \frac{e^{i \theta } }{r} (\omega_{n_0}+\omega)\right)
\Gamma \left(1 + \frac{e^{-i \theta } }{r}  \left(\omega_{n_0}^*-\omega\right)\right)
\Gamma \left(1 + \frac{e^{-i \theta } }{r} \left(\omega+\omega_{n_0}^*\right)\right)
}{\left| \Gamma \left(1+ \frac{e^{i \theta }
   \omega_{n_0}}{r}\right)\right| ^4}, \\
 & \text{On-axis:}~
 \prod_{n = n_0 + 1}^\infty \left( 1 - \frac{\omega^2}{\omega_n^2} \right)^{-1} = \Gamma \left(1+ \frac{ i \left( \omega_{n_0} - \omega \right) }{r} \right) \Gamma \left(1 + \frac{i \left(  \omega_{n_0} + \omega \right) }{r}\right) \Gamma \left(1 + \frac{i \omega_{n_0}}{r}  \right)^{-2}.
 \end{split}
\end{equation}
}}
These tail corrections play a significant role in the high precision matching demonstrated in \fig{fig:RNAdS2pt}.
{\flushleft{\bf{A proof of the thermal product formula:}}}
In the above discussion, we have numerically demonstrated the validity of the thermal product formula \eqref{RNprodform} for the shear modes (see \fig{fig:RNAdS}).
This tremendous numerical agreement begs a direct proof of the validity of this formula.
Such a proof immediately follows from the discussion in \cite{Dodelson:2023vrw}.
The key observation in this regard is that the singularity
structure of the effective potential in our master equations \eqref{chBHmastereqn} is identical to that of a minimally coupled scalar in  similar backgrounds
for which the proof of the formula has been presented in \cite{Dodelson:2023vrw}.
In order to see this point clearly, let us perform a redefinition of the master variables $\Phi_\pm = \sqrt{z} ~\widehat \Phi_\pm$ in \eqref{chBHmastereqn}.
Thereafter we move to tortoise coordinate given by $du = dz/ f(z)$, such that $u \approx - \ln (1-z) / (2(2 - Q^2) \rightarrow \infty$ as we approach the horizon,
while $u \approx z \rightarrow 0$ as we approach the boundary. With these two modifications, \eqref{chBHmastereqn} reduces to
\begin{equation}\label{maseqtor}
 \left( - \partial^2_u  + V_\pm \right) ~\widehat\Phi_\pm = \omega^2 \widehat \Phi_\pm,
\end{equation}
where
\begin{equation}\label{effpot}
 V_\pm \left( u(z) \right) = \frac{1}{12 z^2} f(z) \left(12 k^2 z^2+117 Q^2 z^6+9 -z^4 \left(\pm 8 \mathcal C +33 Q^2+33\right) \right),
\end{equation}
which asymptotically behaves as
\begin{equation}\label{effpotasym}
 V_\pm \left( u(z) \right) =  \begin{cases} \frac{3}{4 z^2} + \dots & , z \rightarrow 0 ~\left(\text{near boundary}\right)
 \\ \frac{1}{6} {\small{\left( 2 - Q^2\right) (84 Q^2 + 12 k^2 - 24 \mp 8 \mathcal C )}} (1-z) + \dots & , z \rightarrow 1 ~\left(\text{near horizon}\right). \end{cases}
\end{equation}
Clearly, \eqref{maseqtor} is structurally identical to the equation of a minimally coupled scalar only differing in the exact form of the potential. In addition,
the asymptotic behaviour of the potential also matches that of the scalar potential both near the boundary and near the horizon.
Due to this observation, the arguments of Appendix C in \cite{Dodelson:2023vrw} is
directly applicable to \eqref{maseqtor} and hence the proof follows. The physical essence of this proof
is analogous to a quantum mechanical problem where the scattering amplitude is expressed in terms of the
bound states of the potential \footnote{We thank S. Pratik Khastgir for valuable discussions on this point.}.
To avoid unnecessary repetition, we refer our readers to \cite{Dodelson:2023vrw} for further details.
One of the important ingredients of this proof is the fact that the two-sided Wightman correlators \eqref{Wdef} for $\Phi_\pm$ do not have
any zeros as a function of $\omega$. Note that, although this is true, but the two point correlators of the currents, like \eqref{TTwhit}, do have zeros.
We will discuss this point in more detail below.

Note that the singularity structure of the equation written in terms of the gauge invariant variables \eqref{chBHfluc} is significantly different from those of the master
variables \eqref{chBHmastereqn}. In fact, if we consider the metric perturbations of an uncharged ($Q=0$) blackhole, its equation can be obtained by setting
$\mathcal B = 0, ~\psi =0$ in \eqref{chBHfluc}. It would appear that the arguments of \cite{Dodelson:2023vrw} are invalid for such an equation since the effective potential
of this equation would have singularities whose location would be determined by the factor $(k^2 f - \omega^2)$. However, if we perform the transformation
\eqref{chBHmastervar} and differentiate the equation accordingly, then we recover an equation similar to \eqref{chBHmastereqn}. This
new form of the equation is structurally similar to that of a scalar field and the arguments of \cite{Dodelson:2023vrw} immediately apply.
This explains the excellent numerical agreement observed in fig.12(b) of \cite{Dodelson:2023vrw} for the shear modes in the uncharged case.
Thus we see that the master variables of \cite{Kodama:2003jz} play a crucial role
in providing us with a straightforward justification of the validity of the thermal product formula.
{\flushleft{\bf{Zeros of the correlation functions:}}}
It is interesting to note that two point functions of the currents, eg. $\langle \mathcal T_{t x_2} \mathcal T_{t x_2} \rangle_\mathcal W$ in \eqref{TTwhit},
have many zeros in the complex $\omega$-plane away from the real axis. This is because we have a sum of two infinite products in the numerator of \eqref{TTwhit}.
This is unlike the correlators corresponding to the master variables which does not have any zeros, and hence the arguments of \cite{Dodelson:2023vrw} straightforwardly
applies to it.

In order to locate the zeros of $\langle \mathcal T_{t x_2} \mathcal T_{t x_2} \rangle_\mathcal W$ in \eqref{TTwhit}, we must find solutions to an equation
which has the schematic form
\begin{equation}\label{zeroscm}
 \prod_{n=1}^\infty \left( \omega^2 - \left( \omega^{(1)}_n \right)^2  \right) + \alpha \prod_{m=1}^\infty \left( \omega^2 - \left( \omega^{(2)}_m \right)^2 \right) = 0.
\end{equation}
where $\{\omega^{(1,2)}_n\}$ are the QNMs corresponding to the master variables $\Phi_\pm$. This structure follows when the inverted product formula of two-sided Wightman functions
for the master variables in \eqref{RNprodform} is substituted into the formula for the correlators such as \eqref{TTwhit}. Also, without loss of generality, we have
$\alpha \leq 1$. Otherwise, we can divide the equation by $\alpha$ to ensure this condition.

In general, it is not very easy to find the zeros of this infinite order polynomial. However, here are some important observations based on the structure of \eqref{zeroscm}.
\begin{enumerate}
\item From the locations of the QNMs in Fig.\ref{fig:CSQNM}, we see that asymptotically
away from the real axis one of the modes in the set $\{\omega^{(1)}_n\}$ comes very close to one mode in the set $\{\omega^{(2)}_n\}$. This implies
that we can approximately assume $\omega^{(1)}_i \approx \omega^{(2)}_j (\equiv \Omega_i,$ say), for $i, j > N$, with $N$ being a large integer. A larger value of $N$ makes the approximation better. With this approximation, the factors $( \omega^2 - \Omega_i^2 ) $ are  common to both the terms in \eqref{zeroscm}, implying that $\omega = \pm \Omega_i$ is a solution.
However, despite this fact, we must note that $\omega = \pm \Omega_i$ is  not the location of a zero for the correlator.
This is because, under our assumption, the correlation function \eqref{TTwhit} will have a
factor of $( \omega^2 - \Omega_i^2 )^2$ in the denominator and one of these factors will cancel with that occuring in the numerator.
We must keep in mind that the assumed coincidence would be exact only in the strict $N \rightarrow \infty$ limit.
\item Generically, there are zeros of the correlation function in the neighbourhood  of the QNM values $\{ \pm \omega^{(1)}_n\}$ (but not coincident), with some exceptions.
This can be deduced as follows. We can expand the polynomial in \eqref{zeroscm}
about one of the QNMs $\{+ \omega^{(1)}_a\}$ to obtain
\begin{equation}\label{aproxzero}
\begin{split}
 \prod_{n=1}^N \left( \omega^2 - \left( \omega^{(1)}_n \right)^2  \right) + \alpha \prod_{m=1}^N \left( \omega^2 - \left( \omega^{(2)}_m \right)^2 \right)
  \approx  A \left( \omega - \omega^{(1)}_a \right) + \alpha B + \mathcal O \left( \omega - \omega^{(1)}_a  \right)^2
 \end{split}
\end{equation}
where $A, B$ are terms which are generated from the infinite products. Notably, for most of the QNMs, $B$ has a factor which is small since there is one QNM from the set  $\{\omega^{(2)}_n\}$
which is near $\{\omega^{(1)}_a\}$. Hence, we can infer that there is a zero at $\omega \approx \omega^{(1)}_a - \Delta,$ where $\Delta = \alpha B / A$ is generally a small correction, when the
large factors cancel between $A$ and $B$. On the other hand, if $\Delta$ is large then this conclusion is not valid, leading to exceptions of this generic behavior. Also,
occurence of zeros away from the QNMs is certainly possible and is not ruled out through this argument.

We were able to verify this conclusion by numerically evaluating the polynomial in \eqref{zeroscm} for complex $\omega$. In our numerical evaluation, we have used the notion of `tail correction' to deal with the infinite order polynomial. The results were not affected substantially even if we used a finite number of terms in the product ($N \approx 15$).

\end{enumerate}

%*****************************************************************************************
\subsection{The extremal limit}\label{ssec:RNext}
%*****************************************************************************************

In this section, we consider the extremal limit of the background black brane by setting $Q= \sqrt{2}$ when its temperature \eqref{Tmudef} vanishes.  This limit has been extensively studied with 4D bulk in several previous works \cite{Edalati:2010hk, Faulkner:2009wj} (also see \cite{Lee:2008xf, Liu:2009dm, Rey:2008zz, Faulkner:2010zz,Faulkner:2010da,Faulkner:2011tm}). In this limit, it is well known that the on-axis QNMs coalesce to form a cut in the complex $\omega$-plane. This leads to a non-analytic behaviour of the two-point functions in this limit and the thermal product formula discussed above is not expected to be valid. However, the low-energy scaling behaviour of the retarded correlators observed in this case has attracted significant attention due to its parallels with quantum criticality related to strange metallic cuprates (see \cite{Hartnoll:2016apf, Iqbal:2011ae} for related reviews).

The master equations in the extremal limit can be obtained by substituting $Q = \sqrt{2}$ in \eqref{chBHmastereqn}. These equations may be solved
to obtain the retarded correlators by computing $G^{\Phi}_\pm$ using \eqref{Gdef}. Here, we record the behaviour of these correlators when $\omega$ is small compared to the scale set by the chemical potential $\mu = \sqrt{6}$. A more detailed analysis of the extremal correlators may reveal a suitably modified product formula in which, as $T \rightarrow 0$, the on-axis poles integrate into a non-analytic function. We leave the investigation of this interesting possibility for future work.

For small frequencies ($\omega \ll \mu$), our discussion here is a review of \cite{Edalati:2010hk} adapted to 5D.
The near-horizon geometry for the extremal AdS black brane \eqref{chBH} is
$\text{AdS}_2 \times  \mathbb R^3$. In this case, the locations of the inner and outer horizons of the charged black brane become degenerate, and the metric function $f(u)$ develops a double zero at the horizon. Consequently, near the horizon as $u \rightarrow 1$, we have $f(u) \approx 12 (1-u)^2 + \dots $.

In order to understand the behaviour of the master variables within the near-horizon geometry, it is convenient to
move to a horizon adapted coordinate $\xi$ which, after a suitable rescaling by $\omega$, reads
\footnote{Note that, our coordinate transformation \eqref{nhcoord} is the leading order term in the $u \rightarrow 1$ limit of the transformation
 $\xi = \frac{\omega u}{12 \left( 1 - u\right)}$.}
\begin{equation}\label{nhcoord}
 \xi = \frac{\omega}{12 \left( 1 - u\right)}.
\end{equation}
Rewriting the extremal master equations \eqref{chBHmastereqn} in this coordinate and retaining leading order terms in the small $\omega$ expansion, we obtain
\begin{equation}\label{NHeq}
 \varphi_\pm''(\xi) + \left( \frac{-k^2-12}{12 \xi ^2} \mp \frac{\sqrt{k^2+6}}{\sqrt{6} \xi ^2}+1 \right) \varphi_\pm(\xi) = 0,
\end{equation}
where $\varphi_\pm = \Phi_\pm (u \rightarrow 1)$ are the master fields in the near-horizon region.
Both these equations in \eqref{NHeq} are identical in form to a massive minimally coupled scalar field in $\text{AdS}_2$ with radius $\ell^2 = 1/12$. The masses of these scalar fields are identified to be
\begin{equation}
 m_\pm^2 = k^2 \pm 2 \sqrt{6} \sqrt{k^2+6}+12.
\end{equation}
This implies that the near horizon master fields are dual to IR CFT operators with effective scaling dimensions $\frac{1}{2} + \nu_\pm$, such that
\begin{equation}\label{nuRN}
 \nu_\pm = \frac{1}{2} \sqrt{1 + \frac{1}{3} \left(k^2 \pm 2 \sqrt{6} \sqrt{k^2+6}+12\right)}.
\end{equation}
The retarded correlator in the IR CFT can be obtained by applying the same method outlined in \ref{sssec:RN2pt} adapted to $\text{AdS}_2$.
After imposition of appropriate in-going boundary conditions
at the horizon ($\xi \rightarrow \infty$), the behaviour of the master fields near the boundary of $\text{AdS}_2$ ($\xi \rightarrow 0$) is given by
\begin{equation}\label{GAdS2}
\begin{split}
& \varphi_\pm (\xi) = \xi^{\frac{1}{2} - \nu_\pm } + \widetilde {\mathfrak{G}}_\pm (k) ~\xi^{\frac{1}{2} + \nu_\pm}  \\
\Rightarrow ~~~& \Phi_\pm(u \rightarrow 1) = (1-u)^{-\frac{1}{2} + \nu_\pm} + \mathfrak{G}_\pm (\omega,k) (1-u)^{-\frac{1}{2} - \nu_\pm},
\end{split}
\end{equation}
where in the second line we have moved back to our $u$ coordinate. Also, we have set the sources to unity ensuring the interpretation of
$\mathfrak{G}_\pm$ as the IR CFT retarded correlators. We can solve \eqref{NHeq} analytically in terms of Bessel functions and after imposition of the aforementioned
boundary conditions, we can compute $\mathfrak{G}_\pm$ exactly (see \cite{Faulkner:2009wj, Edalati:2010hk}). Irrespective of the details,
using \eqref{nhcoord} in \eqref{GAdS2},  we can straightforwardly see that $\mathfrak{G}_\pm \sim \omega^{2 \nu_\pm}$.

By suitably matching \eqref{GAdS2} with solutions of the extremal master equations \eqref{chBHmastereqn} in the region outside $\text{AdS}_2$, the UV CFT correlators
can be obtained order by order, as an expansion in small $\omega$. The conclusion of such an exercise is that Im$G^{\Phi}_\pm$
inherit the scaling behaviour of $\mathfrak{G}_\pm$ at the leading order. Hence the conformal dimensions of the IR CFT operators determine the small $\omega$ scaling exponent of the correlators
\begin{equation}\label{smlom}
 \text{Im} G^{\Phi}_\pm \sim \omega^{2 \nu_\pm}.
\end{equation}
This behaviour is clearly non-analytic for generic non-zero $k$, which corresponds to the emergence of a branch cut on the complex $\omega$ plane for QNMs in the extremal limit. For some specific values of $k$ however, this leading order behaviour of Im$G^{\Phi}_\pm$ can appear to be analytic, but there are subleading logarithmic terms due to which Im$G^{\Phi}_\pm$ retains a non-analytic nature. For instance, for $k=0$, we find $\nu_+ = 3/2, ~ \nu_- = 1/2$ but still Im$G^{\Phi}_\pm$ is non-analytic.
%
%
%
%
%
%
%
%
%
%
%
%
%
%
%
%
%
%
%
%
%
%
%
%
%
%
%
%
%
%
%
%
%
%
%

%*********************************************************************************************************************
\section{Shear Fluctuations in the presence of Chern-Simons term ($\kappa \neq 0$)}\label{sec:kappa}
%*********************************************************************************************************************
%
%
%
We now generalize our results to include the effects of the Chern-Simons term in \eqref{action} which can be tracked throughout the computation with the help of the parameter $\kappa$. We will continue to work with the electrically charged black brane background \eqref{chBH} which remains a solution of \eqref{Eineq} and \eqref{MaxScaleq}, even for non-zero $\kappa$. The linearized equations for the fluctuations \eqref{flucmodes} on this background is given by \eqref{Einalleq} and \eqref{Maxalleq}.

As we have seen previously, the combinations of these fluctuations which are invariant under the residual gauge symmetry is given by
\begin{equation}\label{CSGIvar}
 \begin{split}
 & \mathcal G_1 = k  F_1 + \omega H_1, ~{\mathcal G}_2 = k   F_2 + \omega  H_2 , ~\mathcal B_1 , ~{\mathcal B}_2. \\
 \end{split}
\end{equation}
We have two identical Einstein equations for these gauge-invariant combinations which are given by
\begin{equation}\label{CSEinfluc}
 \begin{split}
 & z \mathcal{G}_i''-3 \mathcal{G}'_i- \frac{\left ( f k^2 - \omega^2 \right) z \mathcal{G}_i}{f^2} + k z^3 \psi' \mathcal{B}'_i   =
   \frac{\omega^2
   f \left(z^4 \psi'^2+ 24 \left( f - 1 \right) \right) }{6 f^2 \left(k^2 f-\omega^2\right)} \left(k z^2 \psi'
   \mathcal{B}_i +\mathcal{G}'_i \right) , \\
 \end{split}
\end{equation}
where $i=1,2$. Note that this equation is identical to the first equation in \eqref{chBHfluc} since the Chern-Simons term being topological does not affect
the Einstein equations. On the other hand, the Maxwell equations definitely receives contribution from the Chern-Simons term and are given by
\begin{equation}\label{CSMaxfluc}
 \begin{split}
 & \qquad \qquad  z f \left( \mathcal{B}''_1 + \frac{k \psi' \mathcal{G}'_1}{k^2
   f-\omega^2}\right)+ \left( \frac{1}{6} z^4 \psi'^2 + 3
   f- 4 \right) \mathcal{B}'_1 \\
 &  \qquad \qquad \qquad \qquad \qquad \qquad + z \left(\omega^2 \left(\frac{z^2
   \psi'^2}{k^2 f-\omega^2}+\frac{1}{f}\right)-k^2\right) \mathcal{B}_1 = i \kappa \left( 8 k z^2 \psi' \right) \mathcal{B}_2,\\
& \qquad \qquad  z f \left( \mathcal{B}''_2 + \frac{k \psi' \mathcal{G}'_2}{k^2
   f-\omega^2}\right)+ \left( \frac{1}{6} z^4 \psi'^2 + 3
   f- 4 \right) \mathcal{B}'_2 \\
 &  \qquad \qquad \qquad \qquad \qquad \qquad + z \left(\omega^2 \left(\frac{z^2
   \psi'^2}{k^2 f-\omega^2}+\frac{1}{f}\right)-k^2\right) \mathcal{B}_2 = - i \kappa \left( 8 k z^2 \psi' \right) \mathcal{B}_1 .
 \end{split}
\end{equation}
Clearly due to the Chern-Simons coupling $\kappa$ the modes $\mathcal G_1, \mathcal B_1$ are now coupled with $\mathcal G_2, \mathcal B_2$.
In order to decouple the equations \eqref{CSEinfluc} and \eqref{CSMaxfluc} we define the master variables \footnote{The
master variables for this system were obtained in \cite{Matsuo:2009xn} where the final form of these variables and their decoupled equations were left implicit. The presentation of the master variables and equations here are more explicit and simplified, which can be readily used for subsequent analysis.}
\begin{equation}\label{CSmastervar}
 \begin{split}
  \Phi^{(1)}_{\pm} =  &  \left( \frac{3 k f \psi'}{2 z^2 \left(\omega^2-k^2 f\right)} \right)
 \left( \mathcal G'_1 + i \mathcal G'_2 \right)
  + \left(\frac{\psi'^2 \left(2 k^2 f+\omega^2\right)}{2 \left( \omega^2-k^2 f \right)}+\frac{6 (1-f)}{z^4}  + A^{(1)}_{\mp}  \right) \left(  \mathcal B_1 + i \mathcal B_2 \right), \\
  \Phi^{(2)}_{\pm} =  &  \left( \frac{3 k f \psi'}{2 z^2 \left(\omega^2-k^2 f\right)} \right) \left(  \mathcal G'_1 - i \mathcal G'_2 \right)
   +  \left(\frac{\psi'^2 \left(2 k^2 f+\omega^2\right)}{2 \left( \omega^2-k^2 f \right)}+\frac{6 (1-f)}{z^4} + A^{(2)}_{\mp}  \right) \left(  \mathcal B_1 - i  \mathcal B_2 \right),  \\
 \end{split}
\end{equation}
where  the constants  $A^{(1,2)}_{\pm}$ are given by
\begin{equation}\label{CSconsts}
 \begin{split}
 & A^{(1)}_{\pm} = \frac{4 \mathcal C^2 \pm 3 \mathcal C (\mathcal C_1 \mp \tilde \kappa k)-36 \tilde \kappa k \left(Q^2+1\right)}{ (4
   \mathcal C \pm 3\mathcal C_1 + 3 \tilde \kappa k)},\\
 & A^{(2)}_{\pm} = \frac{4 \mathcal C^2 \pm 3 \mathcal C (\mathcal C_2 \pm \tilde \kappa k)+36 \tilde \kappa k \left(Q^2+1\right)}{ (4
   \mathcal C \pm 3 \mathcal C_2 - 3 \tilde \kappa k)}, \\
 \end{split}
\end{equation}
where $\tilde \kappa = 16 \sqrt{3} Q \kappa$, $\mathcal C$ is given by \eqref{Cdef} and $\mathcal C_{1,2}$ are given by
\begin{equation}\label{CSCs}
 \begin{split}
 & \mathcal C_1 = \Big( 16 Q^2 \left(3 k^2-\tilde \kappa k+8\right)+(\tilde \kappa k-8)^2+64 Q^4 \Big)^{\frac{1}{2}},\\
 & \mathcal C_2 = \Big( 16 \left(\left(3 k^2+8\right) Q^2+4 Q^4+4\right)+\tilde \kappa k \left(\tilde \kappa k+16 Q^2+16\right) \Big)^{\frac{1}{2}} .\\
 \end{split}
\end{equation}
In terms of these master variables, the equations \eqref{CSMaxfluc} reduces to
\begin{equation}\label{CSMastereq}
 \begin{split}
 \partial_z \left( \frac{f}{z} {\Phi_{\pm}^{(1,2)}}' \right) + \left( \frac{\omega^2- f k^2}{z f} - 12 Q^2 z^3 + 2 z  ~ \mathcal C^{(1,2)}_\pm  \right) {\Phi_{\pm}^{(1,2)}}  = 0,\\
 \end{split}
\end{equation}
where
\begin{equation}\label{Cpmdef}
 \begin{split}
 \mathcal C^{(1)}_\pm &= 2 + 2 Q
   \left(Q - 2 \sqrt{3} \kappa  k\right) \pm \sqrt{Q^2 \left(3 \left(16 \kappa ^2+1\right) k^2+8\right) + 16 \sqrt{3} \kappa  k Q (1+Q^2) +4 Q^4+4},\\
  \mathcal C^{(2)}_\pm &= 2 + 2 Q
   \left(Q + 2 \sqrt{3} \kappa  k\right) \pm \sqrt{Q^2 \left(3 \left(16 \kappa ^2+1\right) k^2+8\right) - 16 \sqrt{3} \kappa  k Q (1+Q^2) +4 Q^4+4}. \\
 \end{split}
\end{equation}
Note that these equations have the same structure as \eqref{chBHmastereqn}, with only difference in the constants $\mathcal C^1_\pm , \mathcal C^2_\pm$ which now have contributions from the Chern-Simons parameter $\kappa$. If we set $\kappa = 0$ we have $\mathcal C^1_\pm = \mathcal C^2_\pm = 2(1+Q^2) \pm  \mathcal C /3$, and \eqref{CSMastereq} reduces to a pair of \eqref{chBHmastereqn} once we use the explicit form of the background \eqref{chBH}.
%
%

%**************************************************************************************
\subsection{QNMs: Dynamical stability and long-lived modes}\label{sssec:CSQNM}
%**************************************************************************************

We now proceed to compute the QNMs for $\kappa \neq 0$. The procedure is identical to that discussed in \ref{subsec:QNM}. At first
we ensure in-going boundary conditions by redefining the master variables
\begin{equation}
{\Phi_{\pm}^{(1,2)}}(z) = f(z)^{- i \frac{\omega}{4\pi T}} ~z ~ {\tilde \Phi^{(1,2)}}_{\pm}(z),
\end{equation}
Near the boundary $z \rightarrow 0$ these variables have the following behaviour
\begin{equation}\label{mastervarbdyCS}
 {\tilde \Phi^{(1,2)}}_{\pm}(z) =  \left( \frac{\mathcal S^{\Phi^{(1,2)}}_{\pm}}{z} + \mathcal O^{\Phi^{(1,2)}}_{\pm} z + \cdots  \right).
\end{equation}
With these redefinitions in place, it is straightforward to find the values of $\omega$ for which \eqref{CSMastereq} admits normalizable in-going solutions with the
the help of the mathematical package `{\it{QNMspectral}}' \cite{Jansen:2017oag} (see footnote \ref{FN:QNMspectral}). Upon analysing the so obtained QNM spectrum for different values of the
Chern-Simons coefficient $\kappa$ and momentum $k$, we are able to record two interesting observations as discussed below.

{\flushleft{\bf{Unstable modes:}}}
% %
%
%
%

\begin{figure}[thb]
\centering
\includegraphics[width = 0.5\textwidth]{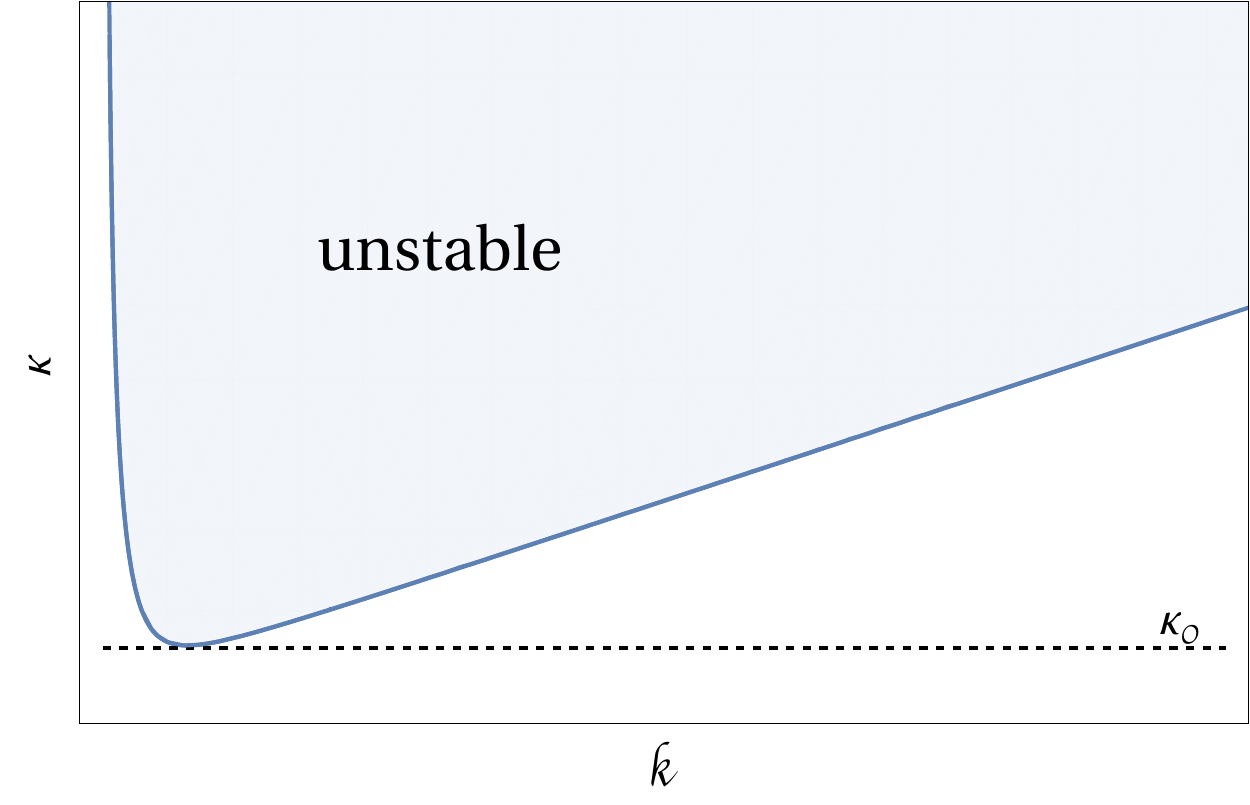}
\caption{Schematic diagram where the shaded region represents the values of Chern-Simons coefficient $\kappa$ and momentum $k$ for which unstable QNMs are observed.
Here $\kappa_0$ is the threshold value of $\kappa$ below which we do not find any instability. When $Q = 1/ \sqrt{2}$, we find $\kappa_0 \approx 0.35$.
}\label{fig:Instab}
%Mathematica file:stability_scheme.nb, evaluation of kappa critical performed in 5D_CSRNADS_Unstable_schematic_plot.nb
\end{figure}
We observe that for some values of $\kappa$ and $k$ one of the four modes in \eqref{CSMastereq} exhibits QNMs with positive imaginary parts. This implies
that the system has dynamical instability for such a value of $\kappa$. For $\kappa > 0$ and $k > 0$, such instabilities are observed in $\Phi^{(2)}_+$, while
QNM spectrum for the rest of the master variables remain stable. For $\kappa < 0$ and $k > 0$, this instability shifts to $\Phi^{(1)}_+$.

\begin{figure}[thb]
\centering
\includegraphics[width = 0.6\textwidth]{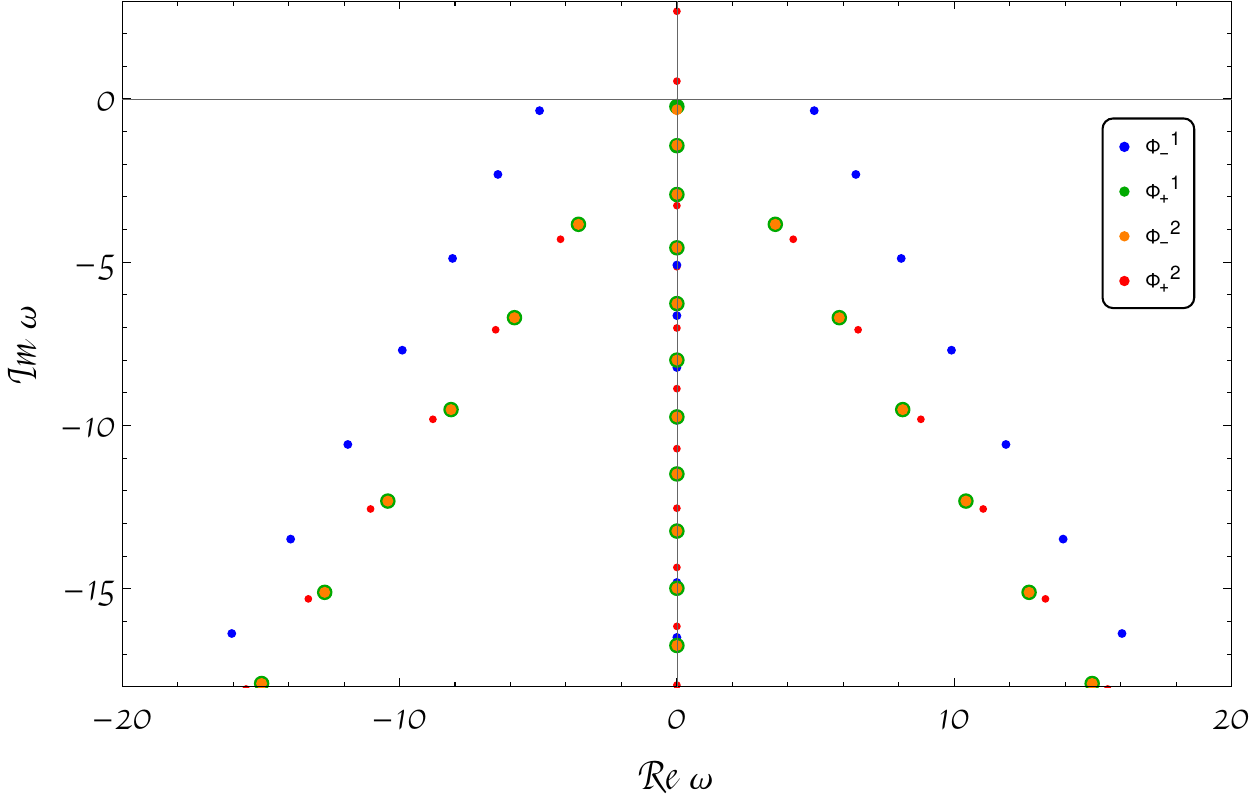}
\caption{Plot of the QNMs in the complex $\omega$ plane when $\kappa = 2$, $Q=1$, $k = 1$. QNMs corresponding
to all the four master variables $\Phi^{(1,2)}_\pm$ have been exhibited. Note that, there are two QNMs corresponding to $\Phi^{(2)}_+$ which develops a positive imaginary part (see the red dots above
the real axis). This signals a dynamical instability of the background electrically charged black brane.
}\label{fig:unstabqnm}
%Mathematica file: 5dcs_RNADS_Unstable_FullPlot.nb
\end{figure}

Although generating the high precision modes is computationally expensive, our investigations suggest that such unstable QNMs arise in a certain region of the $k\kappa$-plane for a given value of background temperature and chemical potential. Sampling over various
values of these parameters, we have schematically plotted the unstable region of the $k\kappa$-plane in \fig{fig:Instab}. Also the unstable QNMs for the specific value of $\kappa=2$ and $k=1$ have been displayed in \fig{fig:unstabqnm} (see the red dots above the real line).

From our numerical experiments it appears that there is an upper bound on $\kappa$ below which all the modes have stable QNMs at all momentum,
while above this threshold value
there is dynamical instability in the system. In fact, for this question, we also considered a scaling limit in which $k \rightarrow 0$
with $\varkappa = k \kappa$ finite. We noticed that there is a critical value $\varkappa_0$ such that this dynamical instability exists only when $\varkappa \geq \varkappa_0$. The value of $\varkappa_0$ depends on $Q$,
for example, for $Q = 1$ we get $\varkappa_0 = 0.6$.

Away from this scaling limit, let us denote the threshold value by $\kappa_0$. We find that $\kappa_0$ depends on the value of $Q$ and it approaches its minimum value  near extremality ($Q \rightarrow \sqrt{2}$). We have also experimented with other values of $Q$, when we find the value of $\kappa_0$ to be higher, possibly
becoming arbitrary large as $Q \rightarrow 0$.
It should be noted that the value of $\kappa = 1 / (4 \sqrt{3}) \approx 0.14$ for $\mathcal N =4$ SYM, which corresponds to a supergravity embedding, lies
just below (but extremely close to) the minimum of $\kappa_0$\footnote{See the discussion in \ref{ssec:CSext} (paragraph below \eqref{extscale}).}.
Hence, $\mathcal N =4$ SYM is not affected by this instability at any value of momentum or temperature.
However, very close to extremality, it is difficult to maintain adequate numerical control on the numerics. It is possible that the supergravity value of
$\kappa$ is somewhat special and coincides with the minimum value of the threshold, approaching it from below in a limiting sense
(see \cite{Horowitz:2024kcx} where the speciality of the supergravity value of $\kappa$ has been emphasized).

{\flushleft{\bf{Long-lived modes:}}}
We also observe that, for a given value of $\kappa$, and for sufficiently large momentum $k$, the imaginary part of some QNMs tends to zero.
Physically, this means that even if the system is dissipative, there are certain excitations of the system whose relaxation time scales
become arbitrarily large compared to other length scales in the system (such as $T$ and $\mu$).
Such anomaly induced long-lived modes are extremely interesting from the boundary theory point of view. Similar observations have been previously reported in several interesting works related to magnetically charged black branes \cite{Waeber:2024ilt, Meiring:2023wwi,Haack:2018ztx} (also see \cite{Ammon:2017ded,Abbasi:2018qzw,Bu:2016oba,Bu:2016vum,Bu:2018psl,Bu:2019mow} for related discussions on transport properties). Hence, it is interesting to observe a similar behaviour even for the simple electrically charged black branes which has been considered here.
\begin{figure}[thb]
\centering
\includegraphics[width = 0.6\textwidth]{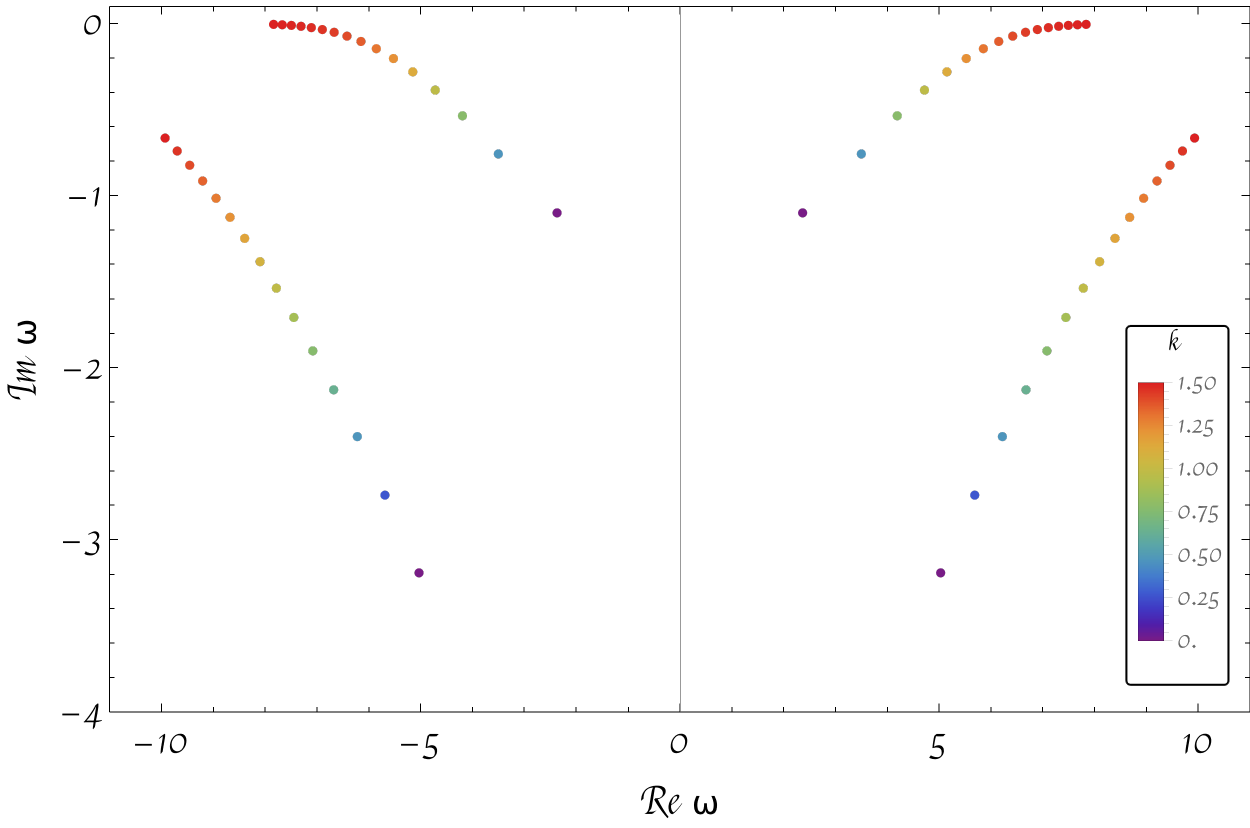}
\caption{In this figure we trace the change in the first two off-axis QNMs corresponding to the
master variable $\Phi^{(1)}_-$ as we vary $k$ from $0$ to $1.5$. Here we have chosen
$Q=1$ and $\kappa = 6$. Clearly, the imaginary part of the first QNM tends to zero as
we increase the value of $k$. This trend shows the existence of long-lived modes
at large momentum for this system.
}\label{fig:longlive}
%Mathematica file: 5dcs_RNADS_LongLived_Modes_Easy_Version.nb
\end{figure}

Firstly, for $\kappa > 0$, such long-lived modes are only observed in the QNM spectrum of $\Phi^{(1)}_-$, which is different
from the master variable in which the instability was observed. For the rest of the master variables the QNMs remain
well separated from the real line for sufficiently large values of momentum.
In order to demonstrate the existence of these long-lived modes, in \fig{fig:longlive} we have plotted the first two off-axis QNMs of $\Phi^{(1)}_-$ with the
lowest  frequencies, i.e. nearest to the real axis. We have tracked these QNMs at $\kappa = 6$ while we increase the value of $k$.
As we can see from the plot, clearly the imaginary part of the first QNM becomes vanishingly small near $k / \mu \sim 1$.
These modes stick to the real axis even when we increase $k$ further, beyond the values shown in the figure. We have chosen
a relatively high value of $\kappa$ in order to demonstrate the effect clearly at relatively smaller values of $k$. Although, for this value of
$\kappa$, we have the aforementioned instability for a large range of momentum and temperature, we believe the phenomenon of long-lived modes is independent
of the existence of the  instability. If we work with values of $\kappa$ below the minimum of the threshold $\kappa_0$, the convergence of the
QNM on the real axis is relatively slower and perhaps takes place at a high value of $k$. For such small values of $\kappa$, we have definitely
observed the trend of convergence but were unable to see complete convergence on the real axis due to numerical limitations.
{\flushleft{\bf{Effective Potential of Master equations: }}}
In order to establish a qualitative understanding of the long-lived modes, it is
instructive\footnote{We thank the anonymous referee for suggesting this analysis of the effective potential.}
to analyse the profile of the effective potentials appearing in the corresponding master equations.
The effective potential can be obtained by performing a manipulation similar to \eqref{maseqtor}. After redefining the master variables
${\Phi_\pm}^{(1,2)} = \sqrt{z} ~ {\widehat{\Phi}_\pm}^{(1,2)}$ and moving to the tortoise coordinate $u$, the master equations \eqref{CSMastereq} reduce to
\begin{equation}
 (- \partial_u^2 + {V_{\pm}}^{(1,2)}){\widehat{\Phi}_\pm}^{(1,2)} = \omega^2 {\widehat{\Phi}_\pm}^{(1,2)},
\end{equation}
where the effective potentials are given by
\begin{equation}\label{cspots}
\begin{split}
V^{(1,2)}_\pm \left(z(u)\right)
= \frac{1}{4 z^2} \left( f(z) \left(z^4 \left(-8 \mathcal C^{(1,2)}_\pm +5 Q^2+5\right)+4 k^2 z^2+39 Q^2 z^6+3\right) \right),
\end{split}
\end{equation}
with $\mathcal C^{(1,2)}_\pm$ given by \eqref{Cpmdef}.

\begin{figure}[h]
\centering
\begin{subfigure}{.5\textwidth}
  \centering
  \includegraphics[width=0.9\linewidth]{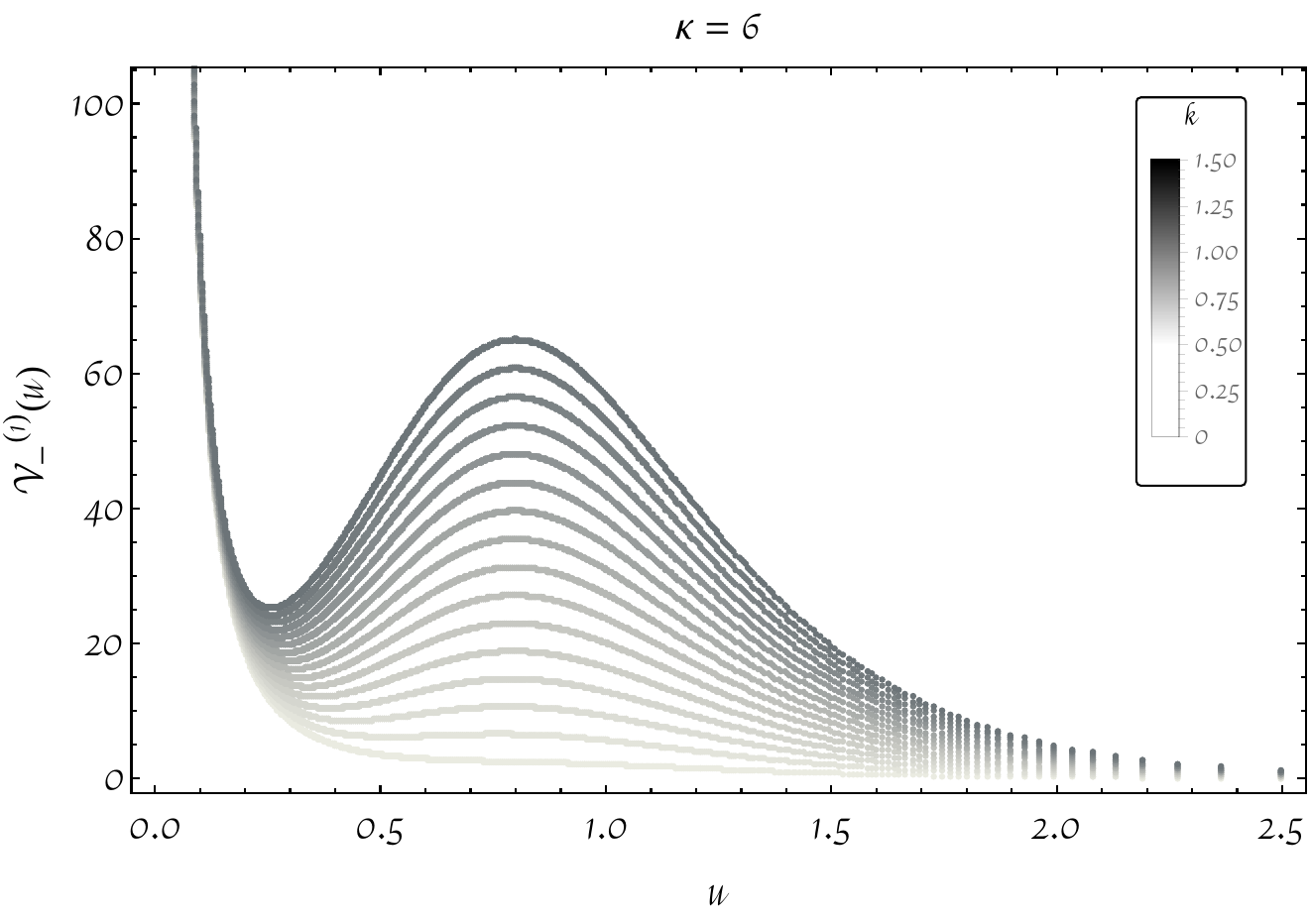}
    \caption{}\label{fig:potkappa}
\end{subfigure}%
\begin{subfigure}{.5\textwidth}
  \centering
  \includegraphics[width=0.9\linewidth]{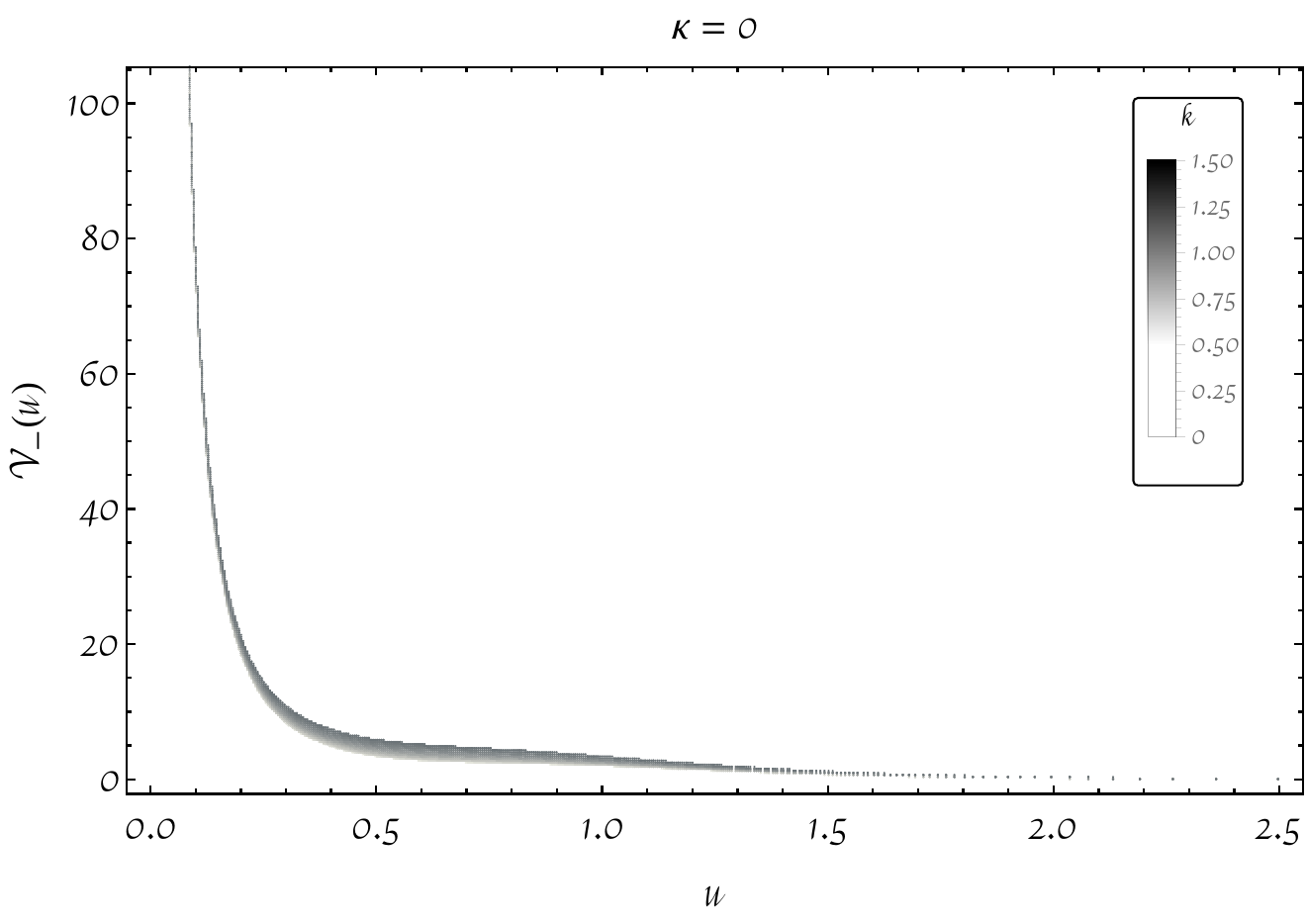}
  \caption{}\label{fig:potRN}
\end{subfigure}
\caption{Plot of the potential ${V_{-}}^{(1)}$ using \eqref{cspots} as a function of the tortoise coordidate $u$. Note that $u \rightarrow 0$ is the AdS-boundary while
$u \rightarrow \infty$ is the horizon of the black brane. In (a), we have plotted the potential for $\kappa = 6$, while in (b), we have set $\kappa =0$. Besides, for both the plots we have
used $Q=1$ and varied $k$ in the range $[0,1.5]$. As we increase $k$, a metastable minima develops for non-zero $\kappa$ indicating the presence of
high momentum long-lived modes in the QNM spectrum. No such minima is observed in (b) for $\kappa =0$.}
\label{fig:PotPlot}
%Mathematica file: DiffEqn_test&_potential_analysis.nb, DiffEqn_test&_potential_analysis_stablePlot.nb
\end{figure}

Note that in Fig.\ref{fig:longlive}, the long-lived modes occur in the QNMs corresponding to the master variable ${\Phi_-}^{(1)}$ when $\kappa =6$. Hence, we have analysed the corresponding potential ${V_{-}}^{(1)}$ in Fig.\ref{fig:PotPlot}. In Fig.\ref{fig:potkappa}, we plot the potential for $\kappa = 6$ and the
same values of $k$ as in Fig.\ref{fig:longlive}. We notice that as we increase $k$,  ${V_{-}}^{(1)}$ develops a metastable minima. This metastable minima is responsible for the
long-lived modes (see section 6 of \cite{Festuccia:2008zx} for futher details). Also, in Fig.\ref{fig:potRN}, we have plotted $V_{-} = {V_{-}}^{(1)}$ at $\kappa =0$, for the same range of $k$.
For vanishing $\kappa$, we do not observe any such metastable minima. This strongly implies that the metastable minima and hence the long-lived modes are a consequence of non-zero $\kappa$, i.e. it is an anomaly induced effect from the boundary point of view.

We have also verified that such metastable minima does not arise in other three potentials ${V_{+}}^{(1)}$ and ${V_{\pm}}^{(2)}$, for the given set of parameters. It is also noteworthy
that the functional form of ${V_{+}}^{(2)}$ is similar to Fig. 6 of \cite{Festuccia:2008zx}, indicating the instability as observed in the spectrum of QNMs in Fig.\ref{fig:unstabqnm}.

%********************************************************************
\subsection{Two-point functions and the product formula} \label{ssec:CS2pt}
%********************************************************************

\begin{figure}[h]
\centering
\begin{subfigure}{.5\textwidth}
  \centering
  \includegraphics[width=0.9\linewidth]{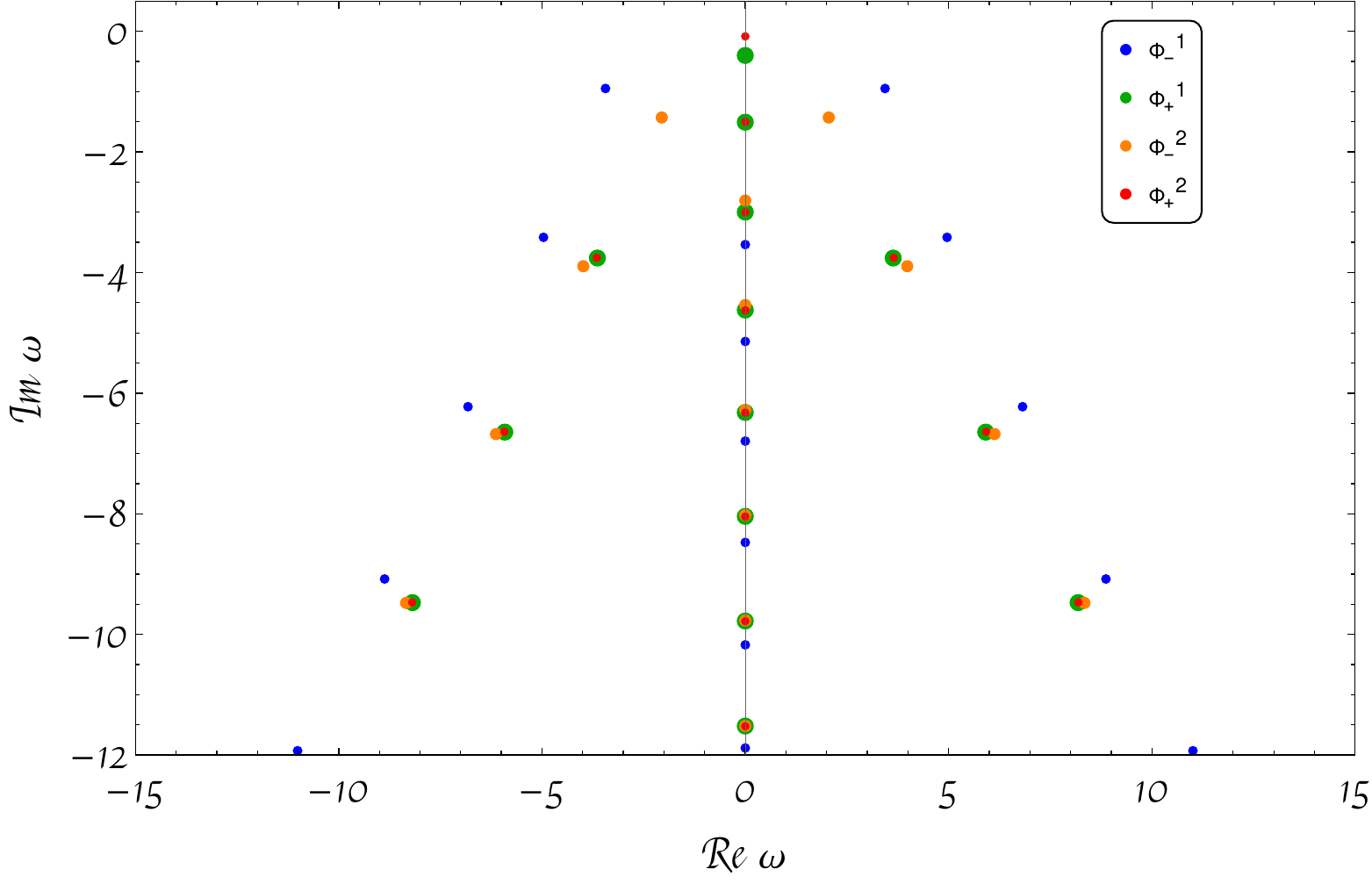}
    \caption{}\label{fig:CSQNM}
\end{subfigure}%
\begin{subfigure}{.5\textwidth}
  \centering
  \includegraphics[width=0.9\linewidth]{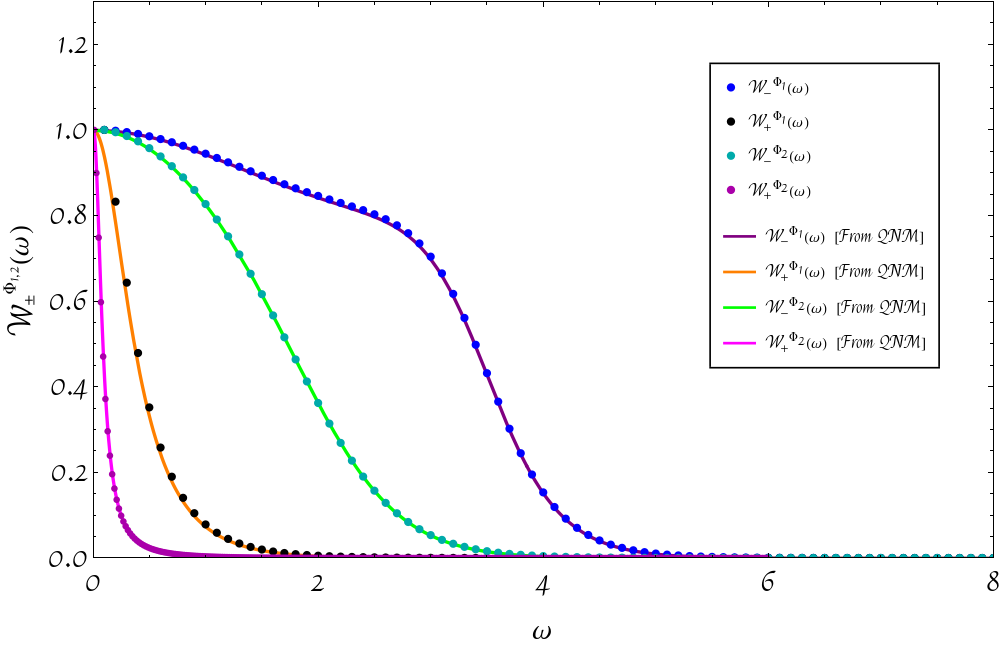}
  \caption{}\label{fig:CS2pt}
\end{subfigure}
\caption{In (a) we have plotted the QNMs arising out of \eqref{CSMastereq} for $\kappa = 0.2$,  $Q= 1$ and $k = 1.5$.
In (b) we plot the two sided Wightman function \eqref{Wdef} for the same set of parameters using two different methods, thus verifying the product formula \eqref{CSprodform}.
These figures are generalizations of their $\kappa = 0$ counterparts shown in \fig{fig:RNAdS}. Also, we adjusted the normalization so that $\zeta^{(1,2)}_\pm = 1$.
In the tail correction formula \eqref{tailcor} we have used $n_0 = 35$, $r \approx 1.75$ for the on-axis poles, and $n_0 = 18$, $r \approx 3.59$, $\theta \approx 0.89$ for the off-axis poles, for all the master variables
$\Phi^{(1,2)}_\pm$.}
\label{fig:CSProdFverify}
%Mathematica file: 5dcs_RNADS_MasterEquations_Sch_Coordinates_kappam6_Q1.nb (mainfile), Final_CS_thermal_product_plot.nb (plot file)
\end{figure}

We shall now proceed to verify the product formula for all two point functions of the relevant components of the energy momentum tensor and charge current.
Following the discussion in \ref{sssec:RN2pt}, the first step is to numerically solve the differential equation \eqref{CSMastereq} with in-going boundary conditions and hence use the ratio of the corresponding normalizable and non-normalizable modes to compute the respective retarded and Wightman correlators
\begin{equation} \label{CSGWdef}
 G^{\Phi^{(1,2)}}_{\pm} (\omega, k) = \frac{\mathcal O^{\Phi^{(1,2)}}_{\pm}(\omega, k)}{\mathcal S^{\Phi^{(1,2)}}_{\pm} (\omega, k)},
 ~~\mathcal W^{\Phi^{(1,2)}}_{\pm}(\omega, k) = \frac{\text{Im} ~G^{\Phi^{(1,2)}}_{\pm} (\omega,k)}{\sinh \left( \omega \beta /2\right)}.
\end{equation}
The Wightman correlators so computed can be matched with the following product formula involving QNMs
\begin{equation}\label{CSprodform}
 \mathcal W^{\Phi^{(1,2)}}_{\pm} = {\zeta_\pm^{(1,2)}} \underbrace{\prod_{n} \left( 1 - \frac{\omega^2}{\omega_n^2} \right)^{-1} \left( 1 - \frac{\omega^2}{{\omega^\star}_n^2} \right)^{-1}}_{\text{(off-axis poles: single line)}}      \underbrace{\prod_{n} \left( 1 - \frac{\omega^2}{\omega_n^2} \right)^{-1}}_{\text{(on-axis poles)}}
\end{equation}
The overall constants remain undetermined ${\zeta^{(1,2)}}$ but they can be computed by fitting the product formula with the result obtained directly by solving the differential equations. From our numerical experiments with different values of $\kappa$ and $k$, we find that the values of ${\zeta_\pm^{(1,2)}}$ depend on both these parameters as well as on $Q$. We have verified that $\mathcal W^{\Phi^{(1,2)}}_{\pm} / {\zeta_\pm^{(1,2)}}$ computed directly from differential equations matches with \eqref{CSGWdef} for various values of $\kappa$. In \fig{fig:CSProdFverify}, we have demonstrated this match for $\kappa = 0.2$,  $Q= 1$ and $k = 1.5$. Note that this value of $\kappa$ is below the threshold value in \fig{fig:Instab} for which we expect the charged black brane background to be stable for all values of momentum.
Also, we have appropriately used the tail correction formula \eqref{tailcor} to obtain this high precision numerical agreement.

We can now express the retarded correlators of the energy momentum tensor and the R-charge current in terms of those corresponding to the master variables.
We derive these relations following the standard AdS-CFT dictionary which is summarized in \App{App:2pt}. The explicit form of these expressions
are somewhat cumbersome (see \eqref{CSallretcor}). However all the retarded correlators have the following general structural form
\begin{equation}\label{CScorelform}
 \begin{split}
 & \langle \mathscr T_1 \mathscr T_2 \rangle = \mathscr C_+^1  G^{\Phi_1}_+
 + \mathscr C_+^2  G^{\Phi_2}_+ + \mathscr C_-^1  G^{\Phi_1}_- + \mathscr C_-^2  G^{\Phi_2}_-,
 \end{split}
\end{equation}
where $\mathscr T_i$ are the different components of the energy momentum and charge currents, $ G^{\Phi_{1,2}}_\pm$ are the retarded correlators corresponding to the master variables defined in \eqref{CSGWdef}, while $\mathscr C_{\pm}^{1,2}$ are specific functions of $\omega, k$ and $\kappa$ which are completely determined by the holographic dictionary.

As discussed in \ref{ssec:Prodform}, we can use \eqref{CScorelform} to write down a product formula for the corresponding two-sided Wightman functions
\begin{equation}\label{CSwform}
 \begin{split}
 & \langle \mathscr T_1 \mathscr T_2 \rangle_{\mathcal W} = \sum_{i=1}^4 \mathscr C_i \mathcal W_i,
 \end{split}
\end{equation}
where in the sum, $\mathscr C_i$ runs over $\mathscr C_{\pm}^{1,2}$ and $\mathcal W_i$ runs over $\mathcal W^{\Phi_{1,2}}_\pm$ defined in \eqref{CSGWdef}. Now,
$\mathcal W^{\Phi_{1,2}}_\pm$ are expressed in terms of the product formula \eqref{CSprodform} so that \eqref{CSwform} reduces to the schematic form
\begin{equation}\label{CSprodF}
\begin{split}
 \langle \mathscr T_1 \mathscr T_2 \rangle_{\mathcal W} = \sum_{i=1}^4 \frac{ \mathscr  C_i \zeta_i}{\prod \left( 1 - \omega^2 / ( \omega^i_n )^2 \right)},
\end{split}
\end{equation}
with $\zeta_i = \{ \zeta^{(1,2)}_\pm \}$ being the four undetermined constants in \eqref{CSprodform}. As discussed previously in \ref{ssec:Prodform}, we can find the values of these constants by appropriately fitting the correlation functions obtained by directly solving the differential equations. It is tempting to
wonder if the ratio of $\zeta_i$ are also expressed in terms of the QNMs such that we can have a completely independent formula
in terms of the QNMs. However, from our limited numerical experiments we unable to find any such connection.

\begin{figure}[thb]
\centering
\includegraphics[width = 0.6\textwidth]{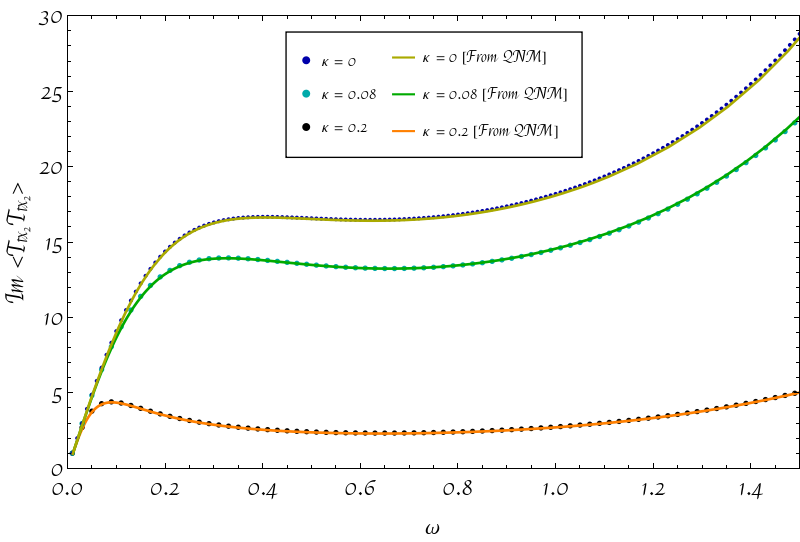}
\caption{Plot of the imaginary part of the correlator $\langle \mathcal T_{t x_2} \mathcal T_{t x_2} \rangle$ using \eqref{CSallretcor}
for different values of $\kappa$ at $Q = 1$ and $k =1.5$. For the dotted lines, we have used numerical solutions of \eqref{CSMastereq} determining $G^{(1,2)}_\pm$ directly. For the solid lines, we have used the product formula \eqref{CSprodform}.
We have determined the constants $\zeta^{(1,2)}_\pm$ using the corresponding matching for the master variables.
We have used,
$\zeta_+ \approx 0.84 \times 10^{-4}$ and $\zeta_- \approx 0.14$ for $\kappa = 0$. While for $\kappa = \{0.08 , 0.2\} $ we get $\zeta^{(1)}_- \approx \{ 0.45 , 0.17 \} \times 10^{-4} $, $\zeta^{(1)}_+ \approx \{0.10, 0.78\} $, $\zeta^{(2)}_- \approx \{1.63 , 5.44  \}\times 10^{-4} $ and $\zeta^{(2)}_+ \approx \{ 0.21, 1.73\}$ respectively. Finally, we have displayed the plots after performing an overall normalization using the values of the correlators at $\omega = 0.01$.
}\label{fig:TTplot}
%Mathematica file:All files in folder: ``xtxt-Correlator_Fin''; Plot:RNAdS_Final_xtxtCorrelator_Plot_Q1_keq 1P5.nb
\end{figure}

Following this strategy, we can write down a product formula for all the correlators in \eqref{CSallretcor}.
In order to demonstrate an example, we plot the imaginary part of the retarded correlator $\langle \mathcal T_{t x_2} \mathcal T_{t x_2} \rangle$ and compare it with the corresponding quantity when $\kappa =0$ in \fig{fig:TTplot}.
Similar plots can be obtained for all the correlators in \eqref{CSallretcor}.
%

%********************************************************************
\subsection{The extremal limit}\label{ssec:CSext}
%********************************************************************

Before concluding, let us consider the extremal limit of the background $Q \rightarrow \sqrt{2}$ and repeat the analysis of \ref{ssec:RNext} for non-zero $\kappa$.
Again, here we are focused on regime where $\omega \ll \mu = \sqrt{6}$. Moving to near horizon coordinates \eqref{nhcoord} and retaining leading order terms in the small $\omega$ expansion, \eqref{CSMastereq} reduces to
\begin{equation}\label{NHeqkappa}
\begin{split}
 {\varphi^{(1,2)}_\pm}''(\xi) + \left( -\frac{\ell^2 m_{(1,2)  \pm}^2 }{\xi^2}+1 \right) \varphi^{(1,2)}_\pm(\xi) = 0,
\end{split}
\end{equation}
where $\varphi^{(1,2)}_\pm = \Phi^{(1,2)}_\pm (u \rightarrow 1)$ represent the master fields in the near horizon region and $\ell^2 = 1/12$ gives the
 AdS$_2$ radius.
These equations again resemble the equation for a massive scalar in AdS$_2$ with masses
\begin{equation}\label{NHmass}
\begin{split}
m^2_{(1)\pm} = ~ & k^2 - 8 \sqrt{6} \kappa  k+12 \mp 2 \sqrt{6} \sqrt{k \left(-8 \sqrt{6} \kappa +16 \kappa ^2 k+k\right)+6}, \\
m^2_{(2)\pm} = ~ & k^2 + 8 \sqrt{6} \kappa  k+12  \mp 2 \sqrt{6} \sqrt{k \left(8 \sqrt{6} \kappa +16 \kappa ^2 k+k\right)+6}.
\end{split}
\end{equation}
The scaling dimension of the corresponding IR operators is given by $1/2 + \nu^{(1,2)}_\pm$ where

\begin{equation}\label{NHnu}
\begin{split}
\nu^{(1)}_\pm = ~  \frac{1}{2} \left( \frac{1}{3} m^2_{(1)\pm} + 1 \right)^{\frac{1}{2}}, ~~
\nu^{(2)}_\pm = ~  \frac{1}{2} \left( \frac{1}{3} m^2_{(2)\pm} + 1 \right)^{\frac{1}{2}}.
\end{split}
\end{equation}
Finally, following the arguments in \cite{Edalati:2010hk, Faulkner:2009wj}, which have been briefly reviewed in \ref{ssec:RNext}, we conclude that the extremal correlators have the following scaling behaviour
for small $\omega$
\begin{equation}\label{extscale}
 \text{Im} G^{\Phi^{(1,2)}}_\pm \sim \omega^{2 \nu^{(1,2)}_\pm}.
\end{equation}
It is interesting to observe how a non-zero $\kappa$ has modified the exponents compared to \eqref{nuRN}.

It may be noted that for certain values of $\kappa > 0$ and $k$, $m_{(2)+}$ goes below the Breitenlohner-Freedman (BF) bound of $AdS_2$ leading
to an instability of the IR near horizon region.
The rest of the three masses in \eqref{NHmass} remain above the BF bound for all $\kappa > 0$ and $k$. Curiously, this region of instability in the $k\kappa$-plane
schematically matches with the region of instability inferred from the QNM spectrum at finite temperature (see \fig{fig:Instab}). The minimum value of
the threshold $\kappa_0$ may be estimated from the analytical expression of $m_{(2)+}$ in \eqref{NHmass}.

Note that when we set $\kappa = 1 / (4 \sqrt{3}) \approx 0.144$ which corresponds to the supergravity value, we find that
$\ell^2 m^2_{(2)+} \geq - 1/4$ for all values of $k$. This suggests that $\mathcal N = 4$ SYM does not suffer from any of this near horizon instability.
However, if we set $\kappa = 0.145$ in \eqref{NHmass}, we find that $\ell^2 m^2_{(2)+} < - 1/4$ for a small range of momentum near $k = 3.63$.
This shows that the supergravity value of $\kappa$ is extremely close to the minimum value of
the threshold $\kappa_0$. Given that there is a small $\omega$ approximation involved  in deriving \eqref{NHeqkappa}, it is possible that
$\mathcal N=4$ SYM may saturate the IR instability bound in the extremal limit.

%
%
%
%
%
%
%
%
%
%
%
%
%
%
%
%
%
%
%
%
%
%
%
%
%
%
%
%
%

%****************************************
\section{Discussions}\label{sec:disco}
%****************************************
%

In this paper, we have investigated the validity of the thermal product hypothesis of \cite{Dodelson:2023vrw} for situations which involve operator mixing. In such a situation, we find that the two sided Wightman function, viewed as a function of frequency or excitation energy, can be written as a sum of products like \eqref{curprodsch} rather than a single product like \eqref{schprodform}. Although exclusive holographic tools were used to reveal this simplified form of the product formula, it has been conjectured in \cite{Dodelson:2023vrw} that such a structure may be universal, even being applicable to large N chaotic thermal systems without any (classical) holographic dual. However, when searching for such a structure in a thermal two-point correlator in a more general setting, we should rather look for a sum-of-product since mixing of operators is very common in general CFTs. The right basis where the operators decouple, which would be equivalent to our master variables, may be very difficult to ascertain entirely from the field theory point of view, especially when it is strongly coupled.  This difficulty would imply that the sum-of-product form would perhaps be more generically observed. Also, it should be noted that in this sum-of-product form \eqref{curprodsch}, there may be zeros in the numerator of the correlators but that does not prohibit them from having a simplified structure.

A somewhat unsatisfying feature of our result \eqref{curprodsch}  has been the presence of the undetermined factors $\zeta_i$. These factors
depend on momentum and other parameters just like the QNMs but they are seemingly not determined by them. Due to their dependence on $k$ they are not mere
normalization of the operators in question. Their behaviour can be retrieved from some prior knowledge of low $\omega$ hydrodynamics.
It would be interesting to explore if these factors can also be expressed entirely in terms of the QNMs, apart from  overall normalizations of the operators. If it is possible, then the entire thermal correlator can be expressed in terms of the QNMs. To this end, it may be worthwhile to seek additional physical input from more general
thermal CFTs and large N chaotic systems.

Another observation which may lead to new and interesting results is related to the extremal limit. From the numerical study of the QNM spectrum in the near extremal limit, it is clear that as we reduce the temperature, distance between the on-axis modes vanishes, while the off-axis modes remain well separated. Finally
the on-axis poles fuse together into a cut, which is reflected in the non-analytic behaviour of the extremal two-point functions. Given the product
formula, it may be possible to integrate the poles in this limit to find an exact form of these non-analyticities. Some of the recent discussions in the
literature on the extremal limit \cite{Horowitz:2022mly, Horowitz:2024kcx, Kolanowski:2024zrq} may be useful in this context.

One important aspect of our study is the inclusion of the Chern-Simons term with a tunable coefficient $\kappa$, which also represents
the 't Hooft anomaly of a global $U(1)$ symmetry of the boundary theory. The current associated with this symmetry is conserved unless there are background non-dynamical electromagnetic fields involving the gauge field coupled to this current. When the hydrodynamics of this system was studied within the framework
of the fluid-gravity correspondence, $\kappa$ was found to be associated with novel transport properties \cite{Banerjee:2008th,Erdmenger:2008rm}. Subsequently, these new transport properties were understood to be universal characteristics of theories with global anomalies \cite{Son:2009tf}. Similar generalization may be possible with some of our results related to $\kappa$. In particular, the dependence of the scaling coefficient of the extremal correlators on $\kappa$  in \eqref{extscale} is quite intriguing.

One of the most striking and rather unexpected revelation in our study has been the dynamical instability of the electrically charged black brane due to
presence of the Chern-Simons term. The end-point of this instability is unclear to us. One possibility is that this instability would lead to some boosted version of the black brane. From the boundary theory point of view, this means that the thermal state with a constant chemical potential (or charge density) is unstable, if
the anomaly coefficient exceeds a certain threshold value. A slight perturbation of this state would perhaps lead to flow of currents, eventually settling down
into a different thermal state. Curiously, in the extremal limit, we find that this instability arises from the IR near-horizon region, which may serve as a clue for this question. It would be interesting to answer this question conclusively in a future investigation. Note that the value of $\kappa$ for $\mathcal N = 4$ SYM appears to lie below the threshold value, for any temperature (see the last paragraph in \ref{ssec:CSext}). Consequently, this instability is entirely absent in $\mathcal N = 4$ SYM.

Relatedly, the QNM spectrum also tells us that for a given value of $\kappa$ as we increase the momentum $k$, the imaginary parts of some of the low frequency QNMs
tends to zero. This implies that there are arbitrarily long-lived modes in this dissipative system. Note that, in our equations (for example, see \eqref{CSMaxfluc}), $\kappa$ always appears multiplied by the momentum. Hence, large momentum in some sense exacerbates the effect of $\kappa$. Therefore, we expect the presence of such long-lived modes at high enough momentum at all values of $\kappa$. From our numerical experiments, it appears that the presence of such long-lived modes is unrelated to the instability. Although we have observed that within the unstable region in \fig{fig:Instab}
the vanishing of the imaginary parts takes place at a faster rate as we increase $k$, compared to the stable region. Below the threshold value of $\kappa$ in that plot, i.e. in the region devoid of any instabilities, it was difficult to observe such long-lived modes since we have to move to a very high value of momentum where the numerical precision was difficult to control. However, even in this case, a trend of the low frequency modes moving towards the real axis was certainly observed.

We believe that there are several open avenues of investigation related to the thermal product formula. For instance, in order to understand the entire scope of field theoretic implications, we should investigate the validity of this result in various holographic backgrounds with different types of fluctuations.
It would be interesting to comprehend how CFT data is organized in more general thermal correlators which we have considered here, along the direction of
\cite{Dodelson:2022yvn, Jia:2024zes, Bhatta:2023qcl}. We should also try to understand the validity and ramifications of these results in specific field theoretic models, such as the SYK models discussed in \cite{Dodelson:2023vrw} (also see \cite{Dodelson:2024atp,Dodelson:2025rng}). The thermal product formula is an elegant example where a mathematical property of a differential equation translates to a deep and important physical result. Understanding the full scope and
applicability of this result would definitely be an enlightening endeavour.

%~~~~~~~~~~~~~~~~~~~~~~~~~~~~~~~~~~~~~~~~~~~~~~~~~~~~~~~
\acknowledgments 
%~~~~~~~~~~~~~~~~~~~~~~~~~~~~~~~~~~~~~~~~~~~~~~~~~~~~~~~
%
We would like to thank Sayantani Bhattacharyya, Shankhadeep Chakrabortty, Diptarka Das, Sajal Dhara, Anjan Kar, S. Pratik Khastgir, Nilay Kundu and Rohan Pramanick for several useful discussions. We are very grateful to Diptarka Das, Sayan Kar, S. Pratik Khastgir, and Nilay Kundu for valuable comments on the draft of this manuscript. AS would like to acknowledge the support from
Prime Minister’s Research Fellowship (\href{https://www.pmrf.in/}{PMRF}), offered by the Government of India.

%****************************************
\appendix
%****************************************
%

%**************************************************************************
\section{Holographic dictionary for two-point functions}\label{App:2pt}
%**************************************************************************

In this appendix, we present the details of the holographic dictionary which expresses the retarded correlators of  the boundary currents in terms of the ratio $G^{\Phi^{(1,2)}}_{\pm}$ in \eqref{CSGWdef} obtained by solving the equation for the master variables \eqref{CSMastereq}. We shall contextually summarize the standard holographic dictionary available in several resources (see for example \cite{Balasubramanian:1999re,Son:2002sd,Son:2006em}), and some of our discussion here overlaps with \cite{Matsuo:2009xn}. Note that, throughout this appendix we shall retain a non-zero  $\kappa$ and in the $\kappa \rightarrow 0$ limit, we  recover \eqref{RNallretcor}.

At first, let us consider the equations for the fluctuations \eqref{flucmodes} which directly follow from the Einstein and Maxwell equation. The relevant components of the Einstein equations provide us with two copies of three identical equations which read
\begin{subequations}\label{Einalleq}
\begin{align}
 & f \left( z F_{1,2}'' + z^3 \mathcal B_{1,2}' \psi '-3 F_{1,2}'\right) - z \ k^2  F_{1,2} -z k
   \omega  H_{1,2} = 0, \label{ein2}\\
 & f \left(6 z f H_{1,2}''+H_{1,2}' \left(6 f+z^4 \psi '^2-24\right)\right)+6z  k \omega
   F_{1,2}+6 z \omega ^2 H_{1,2} = 0, \label{ein1} \\
 & z^2 \omega  \mathcal B_{1,2} \psi '+k f H_{1,2}'+\omega  F_{1,2}' = 0\label{Econsteq},
\end{align}
\end{subequations}
where \eqref{Econsteq}  constitute the two constraint equations in this sector.
On the other hand, the $x_2$ and $x_3$ components of the Maxwell equations read
\begin{equation}\label{Maxalleq}
\begin{split}
 & z f \mathcal B_1 ''+z \mathcal B_1 ' f'-f \mathcal B_1 '+\mathcal B_1  \left(\frac{z
   \omega^2}{f}-k^2 z\right)+z F_1' \psi ' = 8 i \kappa  k z^2  \psi ' \mathcal B_2 ,\\
 & z f \mathcal B_2''+z \mathcal B_2' f'-f
   \mathcal B_2'+\mathcal B_2 \left(\frac{z \omega^2}{f}-k^2 z\right)+z F_2' \psi ' = - 8 i \kappa  k z^2 \psi ' \mathcal  B_1  .\\
\end{split}
\end{equation}
It is possible to solve \eqref{Einalleq} and \eqref{Maxalleq} as a series expansion near the boundary $z \rightarrow 0$ which yields the following asymptotic
expansion for the fluctuations
{\small{
\begin{equation}\label{GFexp}
\begin{split}
  F_{1,2} =& ~\mathcal S^{F_{1,2}} -\frac{1}{4} z^2 (k \mathcal S^{\mathcal G_{1,2}}) + z^4 \mathcal O^{F_{1,2}} - \frac{1}{16} \left( z^4 \ln z \right)  \left( k \left(k^2-\omega^2\right) \mathcal S^{\mathcal G_{1,2}}\right) + \dots, \\
  H_{1,2} =& ~\mathcal S^{H_{1,2}} + \frac{1}{4} z^2 (\omega   \mathcal S^{\mathcal G_{1,2}}) + z^4 \mathcal O^{H_{1,2}} + \frac{1}{16} \left( z^4 \ln z  \right)   \left( \omega \left(k^2-\omega^2\right) \mathcal S^{\mathcal G_{1,2}}\right) + \dots,\\
  \mathcal B_1 =&  ~\mathcal S^{\mathcal B_{1}}  + z^2 \mathcal O^{\mathcal B_{1}} + \frac{1}{64} z^4 \left(8 \mathcal O^{\mathcal B_{1}} \left( k^2-\omega^2 \right)  - 3 \mathcal S^{\mathcal B_{1}} \left(k^2-\omega^2\right)^2  - 8 \sqrt{3}  Q k (\mathcal S^{\mathcal G_1} + 16 i \kappa  \mathcal S^{\mathcal B_{2}}) \right)  \\
 & + \frac{1}{2} \left( z^2 \ln z \right) \left( k^2  - \omega^2 \right)\mathcal S^{\mathcal B_{1}} + \frac{1}{16} \left( z^4 \ln z \right)  \left(k^2-\omega^2\right)^2 \mathcal S^{\mathcal B_{1}} + \dots, \\
  \mathcal B_2 =&  \mathcal S^{\mathcal B_{2}}  + z^2 \mathcal O^{\mathcal B_{2}} + \frac{1}{64} z^4 \left(8 \mathcal O^{\mathcal B_{2}} \left( k^2-\omega^2 \right)  - 3 \mathcal S^{\mathcal B_{1}} \left(k^2-\omega^2\right)^2  - 8 \sqrt{3}  Q k (\mathcal S^{\mathcal G_2} - 16 i \kappa  \mathcal S^{\mathcal B_{1}}) \right)  \\
 & + \frac{1}{2} \left( z^2 \ln z \right) \left( k^2  - \omega^2 \right)\mathcal S^{\mathcal B_{2}} + \frac{1}{16} \left( z^4 \ln z \right)   \left(k^2-\omega^2\right)^2 \mathcal S^{\mathcal B_{2}} + \dots,
  \\
\end{split}
\end{equation}
}}
where we have defined $\mathcal S^{\mathcal G_{1,2}} = k \mathcal S^{F_{1,2}} + \omega \mathcal S^{H_{1,2}}$.
We also have the following constraints
\begin{equation}\label{const}
\begin{split}
 & 2 \left( k \ \mathcal O^{H_{1,2}} + \omega \ \mathcal O^{F_{1,2}} \right)  = \sqrt{3} Q \omega \ \mathcal S^{\mathcal B_{1,2}}.  \\
\end{split}
\end{equation}
Comparing this expansion with \eqref{mastervarbdyCS} after using the transformations \eqref{CSmastervar}, we find that the sources are related as
\begin{equation}\label{Srel}
\begin{split}
 & \mathcal S^{\Phi^{(1)}}_{\pm} = \left( 6 \left( 1 + Q^2 \right) + A^{(1)}_{\mp} \right)  \left( S^{\mathcal B_{1}} + i S^{\mathcal B_{2}} \right)
 - \frac{3\sqrt{3}}{2} k Q \left( \mathcal S^{\mathcal G_{1}} + i \mathcal S^{\mathcal G_{2}} \right),\\
 & \mathcal S^{\Phi^{(2)}}_{\pm} = \left( 6 \left( 1 + Q^2 \right) + A^{(2)}_{\mp} \right)  \left( S^{\mathcal B_{1}} - i S^{\mathcal B_{2}} \right)
 - \frac{3\sqrt{3}}{2} k Q \left( \mathcal S^{\mathcal G_{1}} - i \mathcal S^{\mathcal G_{2}} \right),\\
\end{split}
\end{equation}
while the operator expectation values are related as
{\small{
\begin{equation}\label{Orel}
\begin{split}
  \mathcal O^{\Phi^{(1)}}_{\pm} = & \left( 6 \left( 1 + Q^2 \right) + A^{(1)}_{\mp} \right)  \left( O^{\mathcal B_{1}} + i O^{\mathcal B_{2}} \right) \\
 & - \frac{3\sqrt{3}}{16 \omega} k Q \left( 64 \left( \mathcal O^{H_1} + i \mathcal O^{H_2} \right) + \omega \left( k^2 - \omega^2 \right) \left( \mathcal S^{\mathcal G_{1}} + i \mathcal S^{\mathcal G_{2}} \right) \right), \\
  \mathcal O^{\Phi^{(2)}}_{\pm} = & \left( 6 \left( 1 + Q^2 \right) + A^{(2)}_{\mp} \right)  \left( O^{\mathcal B_{1}} - i O^{\mathcal B_{2}} \right) \\
 & - \frac{3\sqrt{3}}{16 \omega} k Q \left( 64 \left( \mathcal O^{H_1} - i \mathcal O^{H_2} \right) + \omega \left( k^2 - \omega^2 \right) \left( \mathcal S^{\mathcal G_{1}} - i \mathcal S^{\mathcal G_{2}} \right) \right), \\
\end{split}
\end{equation}
}}
where $A^{(1,2)}_{\pm}$ are given by \eqref{CSconsts} and we have used the constraint \eqref{const} to eliminate $\mathcal O^{F_{1,2}}$.
Note that the terms proportional to sources in \eqref{Orel} arise from the $z^4 \ln z$ term in \eqref{GFexp} after substitution in \eqref{CSmastervar}. These terms ultimately
contribute to contact terms in the correlation functions, and hence are not relevant for our discussion here.

After dividing \eqref{Orel} by \eqref{Srel} and recalling the definition of $G^{\Phi^{(1,2)}}_{\pm}$ from \eqref{CSGWdef}, we recover four equations for $\{ \mathcal O^{F_{1,2}}, \mathcal O^{{\mathcal B}_{1,2}} \}$. These equations can be solved to express these operator expectation values linearly in terms of the sources $\{ \mathcal S^{H_{1,2}}, \mathcal S^{F_{1,2}}, \mathcal S^{{\mathcal B}_{1,2}} \}$.

Now, let us evaluate the on-shell boundary action to read off the correlation functions.
After addition of appropriate counter terms \cite{Balasubramanian:1999re,Son:2002sd,Matsuo:2009xn} \footnote{Apart from the counterterms we have considered here,
\cite{Matsuo:2009xn} has also incorporated an $F^2$ counterterm proportional to $\ln z$. We have ignored such $\ln z$ counterterms in the on-shell action since they cannot affect our conclusions here.}, the renormalized action \eqref{action} reads
\begin{equation}\label{Renaction}
\begin{split}
 S =& \int_{\text{bulk}} d^5x \sqrt{-g} \left( \mathcal R + 12 - \frac{1}{4} F_{\mu \nu}F^{\mu \nu}
- \frac{\kappa}{3} \epsilon^{\alpha \mu \nu \sigma \rho} A_{\alpha} F_{\mu \nu} F_{\sigma \rho}  \right) \\
 & + 2 \int_{\partial} d^4x \sqrt{-\gamma} K - 6 \int_{\partial} d^4x \sqrt{-\gamma} - \frac{1}{2} \int_{\partial} d^4x \sqrt{-\gamma} R^{(4)},
\end{split}
\end{equation}
where the boundary term in the second line is the Gibbons-Hawking term \footnote{Since the boundary is a surface of constant $z$, the extrinsic curvature $K = - g^{\mu \nu} \nabla_\mu n_\nu $ where $n_\nu$ is the normalized normal vector orthogonal to a constant $z$ surface, i.e. $n_\nu =  \nabla_\nu z /|\nabla_\nu z|$. }, $\gamma_{\mu \nu}$ is the induced metric on a constant $z$ surface and $R^{(4)}$ is the corresponding Ricci scalar.
We now consider the ansatz
\begin{equation} \label{flucform2}
\begin{split}
 g_{\mu \nu} &= g^{(0)}_{\mu \nu}(z) + \delta g_{\mu \nu} (t, x_1, z), \\
 A_\mu &= A^{(0)}_\mu(z) + \delta A_{\mu} (t, x_1, z), \\
\end{split}
\end{equation}
with \eqref{Schmetanza}, \eqref{chBH} as the background, and substitute it into \eqref{Renaction}. The terms linear in fluctuations vanish by the background equation of motion and for computing the two point functions, we focus on the terms quadratic in fluctuations. An appropriate use of the equations of motion on these quadratic terms
leaves us with only boundary terms and the on-shell action reads
\begin{equation}\label{bdyac}
\begin{split}
  S & =  \int_{\partial} d^4x \left( \sum_{i=1,2} \left(   -\frac{1}{2} \sqrt{3} Q \mathcal B_i F_i + \frac{3 }{2 z^4} \left(1 -\frac{1}{\sqrt{f}} \right)F_i^2  + \left(\frac{\psi'^2}{24}-\frac{1}{z^4} + \frac{3 \sqrt{f} }{2 z^4} -\frac{f }{2 z^4} \right) H_i^2 \right. \right. \\
 & \left.  \left. + \frac{f  }{4 z} \mathcal B_i \mathcal \partial_z B_i -\frac{1}{4 z^3} F_i \partial_z F_i+ \frac{f }{4 z^3} H_i \partial_z H_i
  - \frac{1}{8z^2 \sqrt{f}} \left( \partial_{x_1} F_i - \partial_t H_i \right)^2  \right) - \frac{4}{3} \kappa \psi  \mathcal B_1 \partial_{x_1} \mathcal B_2 \right),
\end{split}
\end{equation}
where we have used the notation \eqref{flucmodes} to denote the fluctuations.
Thereafter, substituting the asymptotic expansion \eqref{GFexp} into \eqref{bdyac}, and considering the appropriate Fourier transform, we obtain
\begin{equation}\label{bdyacOS}
\begin{split}
  S = \int dx_2 dx_3 \int d\omega dk   \sum_{i=1,2} \bigg( & \mathcal O^{H_i} (\omega , k)  \mathcal S^{H_i}(-\omega , -k)  \\ & -  \mathcal O^{F_i}(\omega , k) \mathcal S^{F_i}(-\omega , -k)   + \frac{1}{2}  \mathcal O^{\mathcal B_i}(\omega , k) \mathcal S^{\mathcal B_i}(-\omega , -k) \bigg) + \dots.
\end{split}
\end{equation}
Here, all the terms involving inverse powers of $z$ cancel out leaving behind a finite piece in the $z \rightarrow 0$ limit. However, we do find a term proportional to $\ln z$ which we have ignored along with some other terms proportional to the square of the sources. These terms at best contributes to contact terms in the two point functions. Note that the leading order contribution from the Chern-Simons term is proportional to $\ln z$ whose coefficient is also proportional to the square of sources.

The solutions of $\{ \mathcal O^{H_{1,2}}, \mathcal O^{{\mathcal B}_{1,2}} \}$ from \eqref{Srel} and \eqref{Orel} can now substituted in \eqref{bdyacOS}
while $\mathcal O^{F_{1,2}}$ can be substituted using \eqref{const}. With these substitutions, \eqref{bdyacOS} reduces to the schematic form
\begin{equation}\label{bdyacOSsch}
\begin{split}
  S = \int dx_2 dx_3 \int d\omega dk   \sum  \mathcal S_i (\omega , k) \Pi_{ij}(\omega, k) S_j(-\omega , -k) + \dots.
\end{split}
\end{equation}
The retarded two point functions of the corresponding operators are then given by the following prescription \cite{Son:2002sd}
\begin{equation}\label{prescrip}
\begin{split}
  \langle \mathscr T_i \mathscr T_j \rangle =
  \begin{cases}
 - 2 \Pi_{ij}(\omega, k) , ~~ \text{if} ~ i = j,\\
 - \Pi_{ij}(\omega, k), ~~ \text{if} ~ i \neq j.
\end{cases}
\end{split}
\end{equation}
Since the bulk metric and gauge field couples to the boundary energy momentum tensor and charge current, we compute their retarded two point functions by differentiating $\exp \left(- S\right)$ twice with respect to the corresponding sources, and subsequently set all the sources to zero. The two point functions so obtained by this standard procedure are identical to those obtained by the direct application of the prescription \eqref{prescrip}.
After executing this procedure, we determine the retarded correlation functions of currents in terms of those corresponding to the master variables which, after ignoring contact terms, have been summarized
below
{\small{
\begin{subequations}\label{CSallretcor}
\begin{align}
     \langle \mathcal J_{x_2} \mathcal J_{x_2} \rangle =& 2 \left(1+Q^2+2 \sqrt{3}
   \kappa  k Q \right)\frac{(G^{\Phi_1}_{+}-G^{\Phi_1}_{-}) }{\mathcal{C}_2} + 2 \left(1+Q^2-2 \sqrt{3} \kappa  k
   Q\right) \frac{ (G^{\Phi_2}_{+}-G^{\Phi_2}_{-}) }{\mathcal{C}_1} \nonumber \\
   &-\frac{1}{4} \left( G^{\Phi_1}_{-}+G^{\Phi_1}_{+}+G^{\Phi_2}_{-}+G^{\Phi_2}_{+} \right), \\
        \langle \mathcal T_{x_2 t}  \mathcal T_{x_2 t}  \rangle = &  -\frac{k^2}{2}\left(1+Q^2+2 \sqrt{3} \kappa  k Q\right) \frac{ G^{\Phi_1}_{+}-G^{\Phi_1}_{-} }{\mathcal{C}_2} -\frac{k^2}{2}\left(1+Q^2-2 \sqrt{3} \kappa  k Q\right)
        \frac{ G^{\Phi_2}_{+}-G^{\Phi_2}_{-} }{\mathcal{C}_1}\nonumber \\
   &-\frac{k^2}{16} \left( G^{\Phi_1}_{-}+G^{\Phi_1}_{+}+G^{\Phi_2}_{-}+G^{\Phi_2}_{+} \right), \\
  \langle \mathcal J_{x_2} \mathcal T_{x_2 t}  \rangle =& -\frac{\sqrt{3}}{2}  k^2 Q \left(\frac{ (G^{\Phi_1}_{+}-G^{\Phi_1}_{-}) }{\mathcal{C}_2}+\frac{(G^{\Phi_2}_{+}-G^{\Phi_2}_{-}) }{\mathcal{C}_1}\right), \\
   \langle \mathcal J_{x_2} \mathcal T_{x_2 x_1} \rangle =& -\frac{\sqrt{3}}{2}  k Q \omega \left(\frac{ G^{\Phi_1}_{+}-G^{\Phi_1}_{-} }{\mathcal{C}_2}+\frac{G^{\Phi_2}_{+}-G^{\Phi_2}_{-} }{\mathcal{C}_1}\right) = \frac{\omega}{k} \langle \mathcal J_{x_2} \mathcal T_{x_2 t}  \rangle , \\
   \langle \mathcal T_{x_2 t}  \mathcal T_{x_2 x_1} \rangle =&  \frac{\omega}{k} \langle \mathcal T_{x_2 t}  \mathcal T_{x_2 t}  \rangle, ~~
   \langle \mathcal T_{x_2 x_1}  \mathcal T_{x_2 x_1} \rangle = \frac{\omega}{k} \langle \mathcal T_{x_2 t}  \mathcal T_{x_2 x_1} \rangle = \frac{\omega^2}{k^2} \langle \mathcal T_{x_2 t}  \mathcal T_{x_2 t}  \rangle ,
\end{align}
\end{subequations}
}}
where the constants $C$, $\mathcal C_1$ and $\mathcal C_2$ are given by \eqref{Cdef}, \eqref{CSCs} respectively.
Note that in \eqref{CSallretcor} we have reported the correlators where only $x_2$ component of the currents are involved. Similarly, we have observed that the analogous set of correlators with  $x_3$ component
of the currents have expressions identical to \eqref{CSallretcor}, e.g. $\langle \mathcal T_{x_1 x_2} \mathcal T_{x_1 x_2} \rangle = \langle \mathcal T_{x_1 x_3} \mathcal T_{x_1 x_3} \rangle$ and so on. Besides, we have explicitly checked that all the cross-correlators
such as $\langle \mathcal T_{x_1 x_2} \mathcal T_{x_1 x_3} \rangle$, $\langle \mathcal J_{x_2} \mathcal J_{x_3} \rangle$ etc vanish identically. This simplified structure is similar to the correlators at $\kappa =0$ and is a consequence of the rotational invariance of the background thermal state.

Note that the difference between \eqref{CSallretcor} and \eqref{RNallretcor}
is controlled by the Chern-Simons coefficient $\kappa$, interpreted as the 't Hooft anomaly
from the boundary point of view. Hence \eqref{CSallretcor} provides us with an insight on how the correlators
are modified in the presence of such anomalies. In the $\kappa \rightarrow 0$ limit,
we have \footnote{This follows from the specific form of the mixing of gauge invariant variables in the definition of the master variables, see \eqref{CSmastervar}. Note that, if we set $\kappa = 0$ in \eqref{CSmastervar}, we can recover \eqref{chBHmastervar} only after considering a specific linear combination of the variables.}
\begin{equation}\label{kappa0lim}
  G^{\Phi_1}_{\pm} = G^{\Phi_2}_{\pm}  = G^{\Phi}_{\pm}.
\end{equation}
After this identification, once we set $\kappa = 0$ in the coefficients of  \eqref{CSallretcor} they reduce to those reported in \eqref{RNallretcor}.
%
%
%
%
%
%
%
%
%
%
%
%
%
%
%
%
%
%
%
%%%%%%%%%%%%%%%%%%%%%%%%%%%%%
\bibliographystyle{JHEP}
\bibliography{QNM_TPF}
\end{document}